\documentclass[11pt]{article}
\usepackage{amsmath}
\usepackage{amssymb}

\usepackage[english]{babel}

\def\hybrid{\topmargin 0pt      \oddsidemargin 0pt
        \headheight 0pt \headsep 0pt
        \voffset=-0.5cm
        \hoffset=-0.25in
        \textwidth 6.75in
        \textheight 9.5in       
        \marginparwidth 0.0in
        \parskip 5pt plus 1pt   \jot = 1.5ex}
\catcode`\@=11
\def\marginnote#1{}

\newcount\hour
\newcount\minute
\newtoks\amorpm
\hour=\time\divide\hour by60 \minute=\time{\multiply\hour by60
\global\advance\minute by-\hour}
\edef\standardtime{{\ifnum\hour<12 \global\amorpm={am}%
        \else\global\amorpm={pm}\advance\hour by-12 \fi
        \ifnum\hour=0 \hour=12 \fi
        \number\hour:\ifnum\minute<10 0\fi\number\minute\the\amorpm}}
\edef\militarytime{\number\hour:\ifnum\minute<10 0\fi\number\minute}

\def\draftlabel#1{{\@bsphack\if@filesw {\let\thepage\relax
   \xdef\@gtempa{\write\@auxout{\string
      \newlabel{#1}{{\@currentlabel}{\thepage}}}}}\@gtempa
   \if@nobreak \ifvmode\nobreak\fi\fi\fi\@esphack}
        \gdef\@eqnlabel{#1}}
\def\@eqnlabel{}
\def\@vacuum{}
\def\draftmarginnote#1{\marginpar{\raggedright\scriptsize\tt#1}}
\def\draftlabel#1{{\@bsphack\if@filesw {\let\thepage\relax
   \xdef\@gtempa{\write\@auxout{\string
      \newlabel{#1}{{\@currentlabel}{\thepage}}}}}\@gtempa
   \if@nobreak \ifvmode\nobreak\fi\fi\fi\@esphack}
        \gdef\@eqnlabel{#1}}
\def\@eqnlabel{}
\def\@vacuum{}
\def\draftmarginnote#1{\marginpar{\raggedright\scriptsize\tt#1}}

\def\draft{\oddsidemargin -.5truein
        \def\@oddfoot{\sl preliminary draft \hfil
        \rm\thepage\hfil\sl\today\quad\militarytime}
        \let\@evenfoot\@oddfoot \overfullrule 3pt
        \let\label=\draftlabel
        \let\marginnote=\draftmarginnote
   \def\@eqnnum{(\theequation)\rlap{\kern\marginparsep\tt\@eqnlabel}%
\global\let\@eqnlabel\@vacuum}  }

\def\numberbysection{\@addtoreset{equation}{section}
        \def\theequation{\thesection.\arabic{equation}}}

\def\underline#1{\relax\ifmmode\@@underline#1\else
        $\@@underline{\hbox{#1}}$\relax\fi}

\def\titlepage{\@restonecolfalse\if@twocolumn\@restonecoltrue\onecolumn
     \else \newpage \fi \thispagestyle{empty}\c@page\z@
        \def\thefootnote{\fnsymbol{footnote}} }

\def\endtitlepage{\if@restonecol\twocolumn \else  \fi
        \def\thefootnote{\arabic{footnote}}
        \setcounter{footnote}{0}}  

\numberbysection \hybrid

\newcounter{mo}

\newcommand{\tr}{{\rm tr}}

\newcommand{\mL}{{\mathcal L}}
\newcommand{\mM}{{\mathcal M}}

\newcommand{\mH}{{\mathcal H}}

\newcommand{\mO}{{\mathcal O}}

\newcommand{\vf}{\varphi}
\newcommand{\al}{\alpha}
\newcommand{\be}{\beta}

\newcommand{\om}{\omega}
\newcommand{\vth}{\vartheta}
\newcommand{\de}{\delta}

\newcommand{\Mat}{ {\rm Mat}(N,\mathbb C) }

\newcommand{\mC}{\mathbb C}
\newcommand{\mZ}{\mathbb Z}

\newcommand{\mS}{\mathcal S}

\newcommand{\ka}{\kappa}

\newtheorem{predl}{Proposition}[section]

\def\beq{\begin{equation}}
\def\eq{\end{equation}}
\def\p{\partial}

\def\res{\mathop{\hbox{Res}}\limits}

\begin{document}

\setcounter{page}{1}

\

\vspace{-15mm}

\begin{flushright}
\end{flushright}
\vspace{0mm}

\begin{center}
\vspace{-4mm}
%
{\LARGE{Multi-pole extension for elliptic models  }}
 \\ \vspace{4mm}
  {\LARGE{of interacting integrable tops}}

 \vspace{16mm}

 {\Large  {E. Trunina}\,\footnote{Steklov Mathematical Institute of Russian
Academy of Sciences, Gubkina str. 8, 119991, Moscow, Russia;
\\
 Moscow
 Institute of Physics and Technology, Inststitutskii per.  9,
 Dolgoprudny, Moscow region, 141700, Russia;
 e-mail:
 yelizaveta.kupcheva@phystech.edu.}
 \quad\quad\quad
 {A. Zotov}\,\footnote{Steklov Mathematical Institute of Russian
Academy of Sciences, Gubkina str. 8, 119991, Moscow, Russia;
 e-mail: zotov@mi-ras.ru.}
 }
\end{center}

\vspace{5mm}

\begin{abstract}
 We review
and give detailed description for ${\rm gl}_{NM}$ Gaudin models related to holomorphic vector bundles
of rank $NM$ and degree $M$ over elliptic curve with $n$ punctures.
Then we introduce their generalizations constructed by means of $R$-matrices satisfying the associative Yang-Baxter equation.
A natural extension of the obtained models to the Schlesinger systems is given as well.
\end{abstract}

\bigskip

\

\bigskip


{\small
\tableofcontents
}

\newpage


\section{Introduction}
\setcounter{equation}{0}

In this paper we discuss elliptic integrable systems of classical mechanics, which are described by
the ${\rm gl}_{NM}$-valued Lax matrices with spectral parameter $z$ (it is a coordinate on
elliptic curve $\Sigma_\tau$ with moduli $\tau$), and having simple poles at $n$ marked points.
We also restrict ourselves by considering non-relativistic models, which are governed by linear classical $r$-matrix structures
based on either canonical or the Poisson-Lie brackets on the phase space.

{\bf The aim of the paper} is to present full classification for this type integrable systems by summarizing the previously obtained results
and by introducing the most general type model, which includes all the known as particular cases.
The classification scheme is given on the figure below, and the most general model -- the general ${\rm gl}_{NM}$ Gaudin model
is on its top. It is in the box number 1.
  \beq\label{q303}
   \begin{array}{c}
   \hbox{\underline{Classification scheme for (spin) Calogero and Gaudin type models:}}
   \\ \ \\
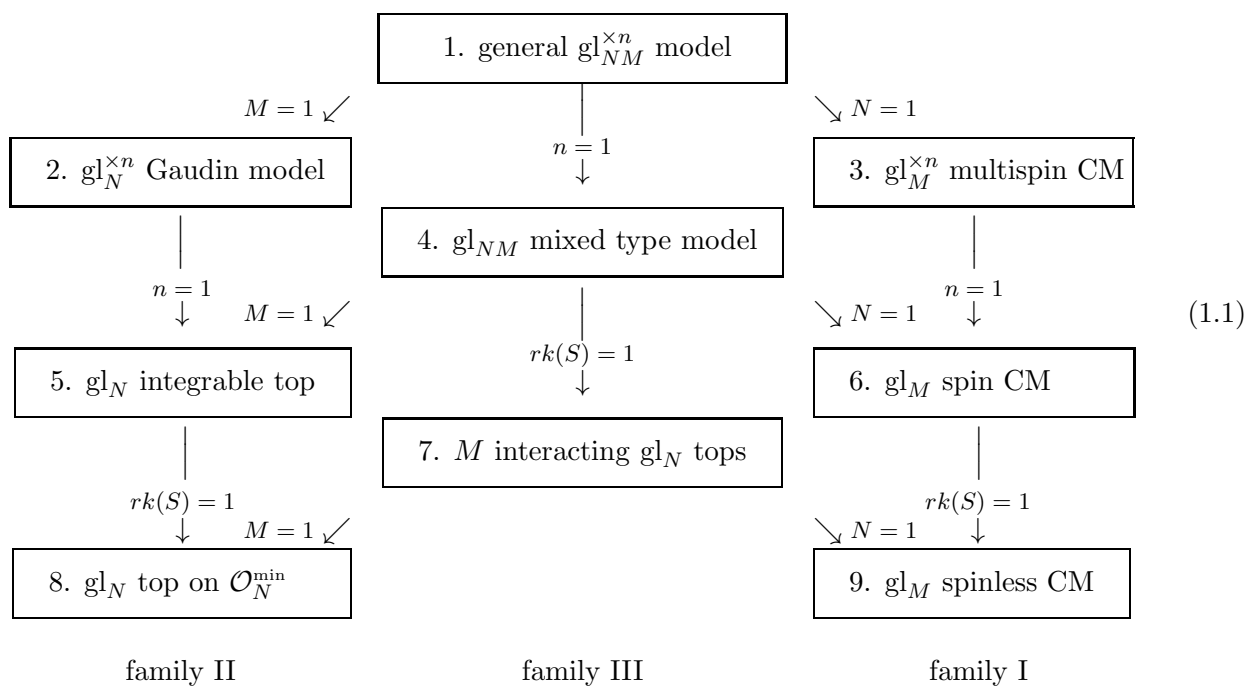
   \begin{array}{ccc}
      & \fbox{ $\phantom{\Big(}$\quad 1. general ${\rm gl}_{NM}^{\times n}$ model\quad $\phantom{\Big(}$} &
   \\
  \hfill$\qquad\qquad\qquad\phantom{\Big(}$ \hbox{\footnotesize{$M=1$}}\swarrow &  \Big| & \searrow
  \hbox{\footnotesize{$N=1$}} $\phantom{\Big(}$\hfill
      \\
    \hfill\fbox{$\phantom{\Big(}$ 2. ${\rm gl}_N^{\times n}$ Gaudin model$\phantom{\Big(}$} &  \stackrel{ \hbox{\footnotesize{$n=1$}} }{\downarrow} &
    \fbox{$\phantom{\Big(}$ 3. ${\rm gl}_M^{\times n}$ multispin CM}\hfill
         \\
   \Big| & \fbox{$\phantom{\Big(}$ 4. ${\rm gl}_{NM}$ mixed type model$\phantom{\Big(}$} & $\phantom{ \hbox{\footnotesize{$n=1$}} }$\qquad\ \ \Big|\hfill
   \\
  $\qquad\qquad\ \phantom{\Big(}$ \stackrel{ \hbox{\footnotesize{$n=1$}} }{\downarrow}\ \, \hfill\hbox{\footnotesize{$M=1$}}\swarrow &  \Big| & \searrow
  \hbox{\footnotesize{$N=1$}}\  $\phantom{\Big(}$\stackrel{ \hbox{\footnotesize{$n=1$}} }{\downarrow}\hfill
      \\
    \hfill\fbox{$\phantom{\Big(}$ 5. ${\rm gl}_N$ integrable top\ $\phantom{\Big(}$} &  $\phantom{\Big(}$
    \stackrel{ \hbox{\footnotesize{$rk(S)=1$}} }{\downarrow} $\phantom{\Big(}$ &
    \fbox{$\phantom{\Big(}$ 6.  ${\rm gl}_M$ spin CM\qquad $\phantom{\Big(}$}\hfill
         \\
   \ \Big| &  \fbox{$\phantom{\Big(}$ 7.  $M$ interacting ${\rm gl}_N$ tops $\phantom{\Big(}$}
  &
  \ \Big|
            \\
 \hfill\stackrel{ \hbox{\footnotesize{$rk(S)=1$}} }{\downarrow} \hbox{\footnotesize{$M=1$}}  \swarrow &
  &
   \searrow
  \hbox{\footnotesize{$N=1$}} \stackrel{ \hbox{\footnotesize{$rk(S)=1$}}}{\downarrow}\hfill
            \\
    \hfill\fbox{$\phantom{\Big(}$ 8. ${\rm gl}_N$ top on $\mO_N^{\hbox{\tiny{min}}}$ \quad $\phantom{\Big(}$} & &
    \fbox{$\phantom{\Big(}$ 9. ${\rm gl}_M$ spinless CM $\phantom{\Big(}$}\hfill
    \\ \ \\
    \hfill\hbox{family II}\qquad\qquad   &  \hbox{family III} & \qquad\qquad  \hbox{family I}\hfill
   \end{array}
      \end{array}
   \eq
In what follows we use matrix valued spin variables $S$. The condition $rk(S)=1$
   means the corresponding minimal coadjoint orbit. We explain it below.

\noindent Let us briefly review the elliptic non-relativistic models from the above scheme. All these models can be roughly
subdivided into two families --  those governed by dynamical classical $r$-matrices (with explicit dependence on the dynamical variables) and those described by non-dynamical $r$-matrices (they depend on spectral parameters but do not depend on the dynamical variables). The first family is presented in the right column, and the second is in the left column. The middle column includes
the intermediate ${\rm gl}_{NM}$ cases, which turn into the first or the second family when $N=1$ or $M=1$ respectively.

{\bf The first family} includes the spinless ${\rm gl}_M$ Calogero-Moser (CM) model \cite{Calogero2} as its basis element (the 9-th box). The Lax representation was introduced by I. Krichever \cite{Kr}, and his ansatz is naturally extended to more complicated models including the most general in the considered class of integrable systems.
The next is the spin generalization of the Calogero-Moser model \cite{GH} (the 6-th box on the scheme).
For the Lax representation we use the approach suggested by E. Billey, J. Avan and O. Babelon \cite{BAB}.
The spin part of the phase space is a symplectic quotient space $\mO//H$ of a coadjoint orbit $\mO$ by the action
of the Cartan subgroup $H\subset {\rm GL}(M,\mC)$. The Poisson bivector and the Lax pair depend on the choice of
gauge fixation conditions entering the Hamiltonian reduction with respect to $H$. To avoid these difficulties one can
describe the model on the unreduced phase space $\mO$ endowed with a simple Poisson-Lie structure.
Then the spinless Calogero-Moser system is described as the unreduced spin CM model on the orbit of minimal dimension.
This is shown by down arrow between cases 6 and 9 on the scheme. The price for this lift (from $\mO//H$ to $\mO$) is appearance
of unwanted terms in the Lax equation, which vanish on the constraints of the Hamiltonian reduction. We explain  more details
in the next Section. Finally, the Gaudin type generalization of the spin CM model was suggested by N. Nekrasov \cite{NN}. It is in the box 3, and is called multispin Calogero-Moser model.
The spin component of the phase space  $\mO^1\times...\times\mO^{n}//H$ includes $n$ coadjoint orbits attached to $n$ marked
points $z_1,...,z_n$ on the punctured elliptic curve $\Sigma_\tau\setminus\{z_1,...,z_n\}$.

{\bf The second family} arises from quantum anisotropic (XYZ) exactly-solvable models \cite{Baxter2} and their quasi-classical description \cite{Skl}. The Lax pairs are constructed using the elliptic Baxter's type $R$-matrix and the Sklyanin's $L$-operator.
The underlying ${\rm gl}_N$ mechanical integrable systems (boxes 5 and 8 on the scheme) are special (elliptic) tops of Euler-Arnold type \cite{LOZ}. Elliptic top can be viewed as a multidimensional generalization of the complexified Euler top in $\mC^3$ with a certain inertia tensor. Its phase space
is a coadjoint orbit $\mO$ of ${\rm GL}(N,\mC)$ Lie group. The Gaudin type generalization was introduced by
A. Reiman and M. Semenov-Tian-Shansky \cite{STS}. The Lax matrix has simple poles at $n$ points on elliptic curve,
and the phase space is given by direct product of $n$ coadjoint orbits $\mO^1\times...\times\mO^{n}$.
In the rational limit it takes the form
  \beq\label{a015}
  \begin{array}{c}
    \displaystyle{
 L(z)=\sum\limits_{k=1}^n\frac{S^k}{z-z_k}\,,\quad S^k\in\mO^k\,.
 }
 \end{array}
 \eq
The models of this type are usually called the Gaudin models, and in physical literature they are often called Hitchin systems.
Let us remark that both titles are somewhat misleading from historical viewpoint. M. Gaudin \cite{Gaudin} studied the quantum models in a special limit. The monodromy matrix of the generalized spin chain on $n$ sites with inhomogeneous parameters $z_k$
can be represented in the form (\ref{a015}), where $S^k$ are quantum operators (representing Lie algebra generators). The Hitchin approach presents integrable systems on the moduli space of Higgs bundles over curves. Originally neither low genus curves nor marked points were considered. This was done later in \cite{NN}, and all the models from the scheme can be indeed described
in the Hitchin framework \cite{NN,LOZ,LOSZ}.

{\bf The middle family} consists of intermediate ${\rm gl}_{NM}$ (the 4-th box) models turning to the first or the second family when
$N=1$ or $M=1$. Originally, they were introduced by A. Polychronakos in his studies of matrix models \cite{Polych}.
Later, the Lax representation for these models was found \cite{ZL} using the Hitchin approach and the M. Atiyah's classification
of vector bundles on elliptic curves \cite{Atiyah}. In the special case, when ${\rm GL}_{NM}$ coadjoint orbit is of minimal dimension, the model can be presented in the form of $M$ interacting ${\rm gl}_N$ elliptic tops (the 7-th box). Finally,
the most general model on the scheme is in the first box. It is the subject of the paper.

Let us also notice that all the families are unified into the so-called symplectic Hecke correspondence \cite{LOZ}. A set of models (from different columns) of the same rank and structure of underlying coadjoint orbits are gauge equivalent since
each model comes from vector bundles of different degrees, and all the bundles can be related by a modification procedure. At the level of the Lax equations the latter means that the corresponding Lax matrices are related by
(singular) gauge transformation degenerated at some point.
Explicit construction of such gauge transformation in the general case  is a complicated problem. But it can be done in some particular cases. For example, the spinless Calogero-Moser model (box 9) is gauge equivalent to the elliptic top
with the orbit of minimal dimension (box 8) when $M=N$. A similar phenomenon in the statistical exactly-solvable models
is known as the IRF-Vertex correspondence \cite{Pasquier} (using this analogy we could call the first family as the IRF type models, and the second family
as the vertex type models).

{\bf In the rational and trigonometric cases} the models of the (spin) Calogero and Gaudin types are classified by the same scheme. However there are more different models due to a variety of possibilities appearing in the limiting procedures applied to a given elliptic one. Instead of specifying all these possibilities (which is a non-trivial task) we use $R$-matrix formulation, when the Lax pair is written in terms of $R$-matrix satisfying the associative Yang-Baxter equation and some set of properties. In the elliptic case the only possible $R$-matrix is the Baxter-Belavin's one. With this $R$-matrix we just reproduce the elliptic models given on the scheme. In the rational and trigonometric limits the models are therefore classified by the same scheme
supplied also by classification of possible (trigonometric or rational) $R$-matrices satisfying the associative Yang-Baxter equation and some additional properties. Such models were previously discussed in \cite{LOZR,LOZ8,GSZ}. Notice that in this way we do not describe all possible trigonometric and rational limits but only those which are represented in the form of spin Calogero and/or Gaudin systems. For example, a class of Toda type models is absent in the $R$-matrix formulation although it can be derived starting from the elliptic models by means of the Inozemtsev limit procedure \cite{Inoz}.

{\bf The paper is organized as follows}. In Section \ref{sect2} we recall constructions of the Lax representations for Calogero-Moser model and its spin generalization. Then we proceed in Section \ref{sect3} to the most general model, and describe
some particular cases in Section \ref{sect4} including the model of interacting tops and the multispin CM system.
The generalized formulation of the obtained results by means of quantum $R$-matrices is given in Section \ref{sect5}.
Finally, the Schlesinger systems are discussed, which are non-autonomous versions of the elliptic integrable models.

\section{Calogero-Moser model and its spin extension}\label{sect2}
\setcounter{equation}{0}

\paragraph{The spinless ${\rm gl}_M$ Calogero-Moser model.} The phase space is $\mC^{2M}$ parameterized by the canonical variables (positions and momenta of particles) with
the canonical Poisson brackets:
  \beq\label{a15}
  \begin{array}{c}
    \displaystyle{
\{p_i,q_j\}=\delta_{ij}\,,\quad \{p_i,p_j\}=\{q_i,q_j\}=0\,.
 }
 \end{array}
 \eq
 The Hamiltonian
  \beq\label{a16}
  \begin{array}{c}
  \displaystyle{
H^{\hbox{\tiny{CM}}}=\sum\limits_{i=1}^M\frac{p_i^2}{2}-\nu^2\sum\limits_{i>j}^M\wp(q_i-q_j)
 }
 \end{array}
 \eq
 describes the pairwise interaction with the potential being the Weierstrass $\wp$-function and the coupling constant $\nu\in\mC$. It provides
 equations of motion
  \beq\label{a14}
  \begin{array}{c}
  \displaystyle{
 {\dot q}_i=p_i\,,\quad  {\ddot q}_i=\nu^2\sum\limits_{k\neq
 i}\wp'(q_{ik})\,.
 }
 \end{array}
 \eq
 The Lax pair with spectral parameter was introduced by I. Krichever \cite{Kr}. It is an explicitly given
pair of $M\times M$ size matrices\footnote{See Appendix for the elliptic functions definitions.}
  \beq\label{a11}
  \begin{array}{c}
  \displaystyle{
L^{\hbox{\tiny{CM}}}_{ij}(z)=(p_i+\nu E_1(z))\delta_{ij}+\nu(1-\delta_{ij})\phi(z,q_{ij})\,,\quad q_{ij}=q_i-q_j\,,
 }
 \end{array}
 \eq
  \beq\label{a12}
  \begin{array}{c}
  \displaystyle{
M^{\hbox{\tiny{CM}}}_{ij}(z)=\nu d_i\delta_{ij}
+\nu(1-\delta_{ij})f(z,q_{ij})\,,\quad d_i=\sum\limits_{k\neq i}
E_2(q_{ik})\,,
 }
 \end{array}
 \eq
which provides the equations of motion (\ref{a14}) through the Lax equation
  \beq\label{q0001}
  \begin{array}{c}
  \displaystyle{
 \dot{L}(z)=[L(z),M(z)]
 }
 \end{array}
 \eq
identically in the spectral parameter $z$. 
  In fact, the Krichever's ansatz for the Lax representation (\ref{a11})-(\ref{a12}) underlies
the Lax pair for the most general model as well.

\paragraph{The spin generalization of the Calogero-Moser model} \cite{GH} (the 6-th box on the scheme).
For the Lax representation we use the approach suggested by E. Billey, J. Avan and O. Babelon \cite{BAB}.
Let us explain the main idea since it is used through out this paper.
Besides the many-body component $\mC^{2M}$  the phase space consists also of the space (as a component in the direct product) parameterized
by variables $S_{ij}$, $i,j=1,...,M$ treated as the classical spin variables. They are naturally arranged into ${\rm gl}(M,\mC)$ valued matrix $S=\sum_{ij}E_{ij}S_{ij}$. The spin component of the phase space is $\mO//H$, where $\mO$ is a coadjoint orbit of
${\rm GL}(M,\mC)$ Lie group, $H\subset {\rm GL}(M,\mC)$ is its Cartan subgroup, and the double factor $//$ means performing the Hamiltonian (or the Poisson) reduction of $\mO$ with respect to adjoint action of $H$. This action (the conjugation $S\rightarrow hSh^{-1}$, $h\in H$ -- diagonal $M\times M$ matrix) provides the moment map constraints
%
  \beq\label{q0003}
  \begin{array}{c}
  \displaystyle{
 S_{ii}=S_{jj}={\rm const}\,,\quad i,j=1,...,M\,.
 }
 \end{array}
 \eq
Being supplied with some gauge fixation conditions $\varsigma_k$, $k=1,...,M$ (fixation of the action $S\rightarrow hSh^{-1}$) they form $2M$ Dirac second class constraints. The Poisson bivector on the reduced phase space $\mO//H$ depends on the choice of the gauge fixing conditions $\varsigma_i$, since they enter the Dirac brackets formula (for a pair of functions $f_{1,2}$ on the reduced space)
  \beq\label{q0002}
  \begin{array}{c}
  \displaystyle{
  \{f_1,f_2\}\Big|_{\rm red}=\Big( \{f_1,f_2\}-\{f_1,\chi\}C^{-1}\{\chi^T,f_2\} \Big)\Big|_{\rm on\ shell}\,,
 }
 \end{array}
 \eq
  where $\chi$ is $2M$ dimensional row $(S_{11},...,S_{MM},\varsigma_1,...,\varsigma_M)$, and $C\in{\rm Mat}_{2M}$ is the matrix
  with elements $C_{kl}=\{\chi_k,\chi_l\}$, $k,l=1,...,2M$. The ''on shell'' means restriction on the constraints.

Instead of dealing with the reduced brackets (\ref{q0002}) which requires some choice of $\varsigma_k$, one can describe
the spin Calogero-Moser model on the unreduced phase space $\mC^{2M}\times \mO$. Then the spin component of the phase space
is equipped with a natural and simple Poisson brackets -- the Poisson-Lie structure on ${\rm gl}^*(M,\mC)$:
  \beq\label{a25}
  \begin{array}{c}
  \displaystyle{
\{S_{ij},S_{kl}\}=-S_{il}\delta_{kj}+S_{kj}\delta_{il}\,.
 }
 \end{array}
 \eq
It remains the same on a coadjoint orbit $\mO$ since the latter is obtained from ${\rm gl}^*(M,\mC)$ by fixation of the Casimir
functions only, i.e. fixation of the eigenvalues of the matrix $S$.
 Following \cite{BAB} introduce the Lax pair
  \beq\label{a24}
  \begin{array}{c}
  \displaystyle{
L^{\hbox{\tiny{spin}}}_{ij}(z)=\delta_{ij}(p_i+S_{ii}E_1(z))+(1-\delta_{ij})S_{ij}\phi(z,q_{ij})\,,
 }
 \end{array}
 \eq
  \beq\label{a36}
  \begin{array}{c}
  \displaystyle{
  M^{\hbox{\tiny{spin}}}(z)_{ij}=(1-\delta_{ij})S_{ij}f(z,q_i-q_j)\,.
 }
 \end{array}
 \eq
The Hamiltonian
  \beq\label{a161}
  \begin{array}{c}
  \displaystyle{
H^{\hbox{\tiny{spin CM}}}=\sum\limits_{i=1}^M\frac{p_i^2}{2}-\sum\limits_{i>j}^M S_{ij}S_{ji}\wp(q_i-q_j)
 }
 \end{array}
 \eq
obtained from $\tr (L^{\hbox{\tiny{spin}}}(z))^2$, and the Poisson brackets (\ref{a15}), (\ref{a25}) provide
equations of motion
  \beq\label{a33}
  \begin{array}{c}
  \displaystyle{
{\dot q}_i=p_i\,,\quad {\ddot q}_i=\sum\limits_{j\neq
i}^MS_{ij}S_{ji}\wp'(q_i-q_j)\,,
 }
 \end{array}
 \eq
  \beq\label{a34}
  \begin{array}{c}
  \displaystyle{
{\dot S}_{ii}=0\,,\quad {\dot S}_{ij}=\sum\limits_{k\neq i,j}^M
S_{ik}S_{kj}(\wp(q_{ik})-\wp(q_{jk}))-(S_{ii}-S_{jj})S_{ij}\wp(q_{ij})\,,\ i\neq j\,.
 }
 \end{array}
 \eq
They are equivalently represented in the form of the Lax equation with additional unwanted term:
  \beq\label{a35}
  \begin{array}{c}
  \displaystyle{
 \dot{L}(z)=[L(z),M(z)]-\sum\limits_{i\neq j}^ME_{ij}\,(S_{ii}-S_{jj})S_{ij}\Big(E_1(z)f(z,q_{ij})+\wp(q_{ij})\phi(z,q_{ij})\Big)\,.
 }
 \end{array}
 \eq
Thus, the spin CM model is not necessarily integrable on the unreduced space, and it becomes integrable on the constraints
(\ref{q0003}) when the unwanted term vanishes. However, equations of motion (\ref{a33})-(\ref{a34}) are
no more valid on the reduced phase space, since the reduction includes not only restriction on-shell the constraints but some more (Dirac) terms coming from the second term in (\ref{q0002}). It is easy to see by considering example of the coadjoint orbit of minimal dimension. In this case $S$ is a rank one matrix $S_{ij}=\xi_i\eta_j$, so that $M-1$ of $M$ eigenvalues of $S$ coincide.
The constraints $S_{ii}=\nu$ being supplied with the gauge fixation $\xi_i=1$ $\forall i$ lead to a trivial spin space after reduction: $S_{ij}=\nu$, so that in this case the spin variables are absent, and we come to the spinless case (\ref{a11}). The Dirac terms result in non-trivial diagonal part of the accompanying matrix (\ref{a12}), while it is zero for the unreduced $M$-matrix (\ref{a36}).
In this way we describe the spinless Calogero-Moser system as the unreduced spin CM model on the orbit of minimal dimension.



\section{Lax pair in the general case}\label{sect3}
\setcounter{equation}{0}

\subsection*{Lax matrix in the general case}
In the general case the Lax matrix has size $NM\times NM$. We represent it in block-matrix form having
$M\times M$ blocks of size $N\times N$ each:
\beq\label{a41}
\begin{array}{c} \displaystyle{
    \mL(z) = \sum_{i,j=1}^M E_{ij} \otimes \mL^{ij}(z) \in {\rm Mat}(NM,\mC)\,, \quad \mL^{ij}(z) \in {\rm Mat}(N, \mC)\,,
}\end{array}
\eq
where $E_{ij}$ is the standard basis in ${\rm Mat}(M,\mC)$. Inside any of $N\times N$ block we use another matrix basis $T_{\al}$:
%
%
%
\beq\label{a971}\begin{array}{c} \displaystyle{
        T_\al = \exp \left( \al_1 \al_2 \frac{\pi i}{N} \right) Q^{\al_1} \Lambda^{\al_2}, \quad \al = (\al_1, \al_2)\in \mZ_N \times \mZ_N, \quad T_0=T_{(0,0)} = 1_N,
}\end{array}\eq
in terms of the generators of non-commutative torus (the finite-dimensional representation of the Heisenberg group)
\beq\label{a972}\begin{array}{c} \displaystyle{
    Q_{jk} = \de_{jk} \exp \left( \frac{2\pi i}{N} k\right), \quad \Lambda_{jk} = \de_{j-k+1 = 0\ \hbox{mod} N}, \quad Q^N = \Lambda^N =1_N\,.
}\end{array}\eq
 The {\em commutation relations} take the form:
\beq\label{Tcond}\begin{array}{c} 
\displaystyle{
    T_\al T_\be = \ka_{\al, \be} T_{\al + \be}, \quad \ka_{\al, \be} = 
    \exp \left( \frac{\pi i}{N}(\al_2 \be_1 - \al_1 \be_2) \right), \quad \ka_{\al, \al + \be} = 
    \ka_{\al, \be}, \quad \ka_{- \al, \be} = \ka_{\be, \al},
}\end{array}\eq
\beq\label{TrT}\begin{array}{c} \displaystyle{
    \hbox{tr} (T_\al T_\be) = N \de_{\al + \be}, \quad \de_\al = \de_{\al_1, 0} \de_{\al_2, 0},
}\end{array}\eq
\beq\label{braketsT}\begin{array}{c} \displaystyle{
    [T_\al, T_\be] = (\ka_{\al, \be} - \ka_{\be, \al}) T_{\al+ \be} = 
    2i \sin \left( \frac{\pi}{N}(\al_2 \be_1 - \al_1 \be_2) \right) T_{\al + \be}.
}\end{array}\eq
The matrix blocks in (\ref{a41}) are of the form:
\beq\label{nerL}
\begin{array}{c}
\displaystyle{
    \mL^{ij}(z) = \de_{ij} \Big( p_i 1_N + \sum_{a=1}^n \mS^{ii,a}_{0,0}\, 1_N\,  E_1(z-z_a) + \sum_{a=1}^n \sum_{\al \neq 0} \mS^{ii,a}_\al T_\al \vf_\al (z-z_a, \om_\al) \Big) +
    }
    \\
    \displaystyle{
    + (1 - \de_{ij}) \sum_{a=1}^n \sum_\al \mS^{ij,a}_\al T_\al \vf_\al (z-z_a, \om_\al + \frac{q_{ij}}{N})\,.
    }
\end{array}
\eq
The index $\al=(\al_1,\al_2)$ take values in $\mZ_N\times\mZ_N$. The sum over $\al\neq 0$ means that we skip
in summation $\al=(0,0)$, which corresponds to $T_{(0,0)}=1_N$ -- the identity matrix, and $\om_\al=0$ (\ref{varphi}).

The Lax matrix (\ref{nerL}) is a natural extension of ${\rm gl}_{NM}$ model to the multi poles case (with $n$ marked points on the elliptic curve $\Sigma_\tau$):
\beq\label{a42}\begin{array}{c} \displaystyle{
    \mS^{ij, a} = \res\limits_{z = z_a} \mL^{ij}(z) \in \Mat\,.
}\end{array}
\eq
When $n=1$ we come back to the mixed type ${\rm gl}_{NM}$ model (case 4 on the scheme).

The origin of the explicit expression for the Lax matrix (\ref{a41}), (\ref{nerL}) is as follows. As was shown
in \cite{ZL,LOSZ} the Lax matrices are classified by the structure of underlying (Higgs) bundles over elliptic curve. The general classification of bundles is known from \cite{Atiyah}. Here we deal with the holomorphic
vector bundle $V$ of degree $M$ and rank $NM$. The Lax matrix is a section of ${\rm End}(V)$ bundle with the following transition functions:
\beq\label{a421}\begin{array}{c} \displaystyle{
    \mL(z+1)=g_1\mL(z) g_1^{-1}\,,\qquad \mL(z+\tau)=g_\tau\mL(z) g_\tau^{-1}\,,
}\end{array}
\eq
where $g_1$ and $g_\tau$ are $NM$ by $NM$ matrices having the following block diagonal structure:
\beq\label{a422}\begin{array}{c} \displaystyle{
     g_1=\bigoplus\limits_{k=1}^M Q^{-1}\,,\qquad
     g_\tau=\bigoplus\limits_{k=1}^M \exp(-2\pi\imath\frac{q_k}{N})\Lambda^{-1}\,.
}\end{array}
\eq
Residues of $\mL(z)$ are fixed as
\beq\label{a423}\begin{array}{c} \displaystyle{
    \res\limits_{z = z_a} \mL(z)=\mS^{a}  \in {\rm Mat}(NM,\mC)\,.
}\end{array}
\eq
Solution of (\ref{a423}) with the quisi-periodic conditions (\ref{a422}) is given by (\ref{a41}), (\ref{nerL}).

The Poisson brackets for momenta and positions of particles are canonical (\ref{a15}).
The set of classical spin variables $\mS^{ij, a}_\al$, $i,j=1,...,M$, $\al\in\mZ_N\times\mZ_N$, $a=1,...,n$ parameterizes $n$ Lie coalgebras
${\rm gl}_{NM}^*$. So that the Poisson brackets are given by the Lie -Poisson structure, which is dual to the
basis $E_{ij}\otimes T_\al$ in ${\rm Mat}(NM,\mC)$. Namely,
\beq \label{a43} \begin{array}{c} \displaystyle{
    \{ \mS^{ij,a}_\al, \mS^{km,b}_\be \} = \frac{\de^{ab}}{N}\Big( \de^{im} \ka_{\al,\be} \mS^{kj,a}_{\al + \be} - \de^{kj} \ka_{\be,\al} \mS^{im,a}_{\al + \be}\Big)\,.
}\end{array}\eq

\subsection*{Hamiltonian description}
The generating function of Hamiltonians appears in the usual way:
\beq\label{a45}
\begin{array}{c}
 \displaystyle{
    \frac{1}{2N}\, \tr \Big(\mL^2(z)\Big) = H_0\! +\! \sum_{a=1}^n H_{1,a} E_1(z-z_a)\! +\! \sum_{a=1}^n H_{2,a} E_2(z-z_a)
    \!+\!\sum\limits_{i=1}^M\sum\limits_{a,b=1}^n\mS_{0,0}^{ii,b}\mS_{0,0}^{ii,a}\rho(z-z_a)\,,
    }
\end{array}\eq
where the last term contains non-double-periodic function in $z$ (\ref{a973}). This term can be removed using additional constraints. We discuss it below.

In order to compute the l.h.s. of (\ref{a45}) we use the property (\ref{TrT}). This yields
\beq\label{TrL2}
\begin{array}{c}
  \displaystyle{
    \frac{1}{2N}\, \tr \Big(\mL^2(z)\Big) = \sum\limits_{i=1}^M\Big[ \frac{p_i^2}{2} +
     \frac{1}{2} \sum\limits_{a,b=1}^n \mS^{ii,a}_{0,0} \mS^{ii,b}_{0,0} E_1(z-z_a) E_1(z-z_b)  +
    }
    \\ \ \\
    \displaystyle{
     + p_i \sum\limits_{a=1}^n \mS^{ii,a}_{0,0} E_1(z-z_a)
     + \frac{1}{2} \sum\limits_{a,b=1}^n \sum_{\al \neq 0} \mS^{ii,a}_\al \mS^{ii,b}_{-\al} \vf_\al ( z-z_a, \om_\al) \vf_{-\al} (z - z_b, \om_{-\al}) \Big] +
    }
    \\ \ \\
    \displaystyle{ + \frac{1}{2} \sum\limits_{i\neq j}^M \sum\limits_{a,b=1}^n \sum_{\al} \mS^{ij,a}_\al \mS^{ji,b}_{-\al} \vf_\al (z-z_a, \om_\al + \frac{q_{ij}}{N}) \vf_{-\al} (z - z_b, \om_{-\al} + \frac{q_{ji}}{N})\,.
    }
\end{array}
\eq
Next, one should use identities (\ref{Ep}), (\ref{diffsign}), (\ref{formulaEf}), which lead to the following answer
for the Hamiltonians from the r.h.s. of (\ref{a45}):
\beq\label{H0}
\begin{array}{c}
\displaystyle{
    H_0 = \sum\limits_{i=1}^M \frac{p_i^2}{2} +
    \frac{1}{2} \sum\limits_{i=1}^M \sum\limits_{a, b}^n \mS^{ii,a}_{0,0} \mS^{ii,b}_{0,0} \rho(z_{ab}) +
     \frac{1}{2} \sum\limits_{i=1}^M \sum\limits_{a,b=1}^n \sum_{\al \neq 0} \mS^{ii,a}_\al \mS^{ii,b}_{-\al} f_\al (z_{ba}, \om_\al) +
    }
     \\ \ \\
     \displaystyle{
     + \frac{1}{2} \sum\limits_{i\neq j}^M \sum\limits_{a,b=1}^n \sum_{\al} \mS^{ij,a}_\al \mS^{ji,b}_{-\al} f_\al (z_{ba}, \om_\al + \frac{q_{ij}}{N})\,,\quad z_{ab}=z_a-z_b\,,\quad \rho(0)=\frac{\vth'''(0)}{3\vth'(0)}\,,
    }
\end{array}
\eq
\beq\label{a46}\begin{array}{c}
 \displaystyle{
        H_{1,a} = \sum\limits_{i=1}^M p_i \mS^{ii,a}_{0,0} + \sum\limits_{i=1}^M \sum\limits_{b: b\neq a}^n \mS^{ii,a}_{0,0} \mS^{ii,b}_{0,0} E_1(z_{ab}) - \sum\limits_{i=1}^M \sum\limits_{b: b\neq a}^n \sum_{\al \neq 0} \mS^{ii,a}_\al \mS^{ii,b}_{-\al} \vf_\al (z_{ba}, \om_\al) -
        }
        \\ \ \\
        \displaystyle{ - \sum\limits_{i,j: i\neq j}^M \sum\limits_{b: b\neq a}^n \sum\limits_{\al} \mS^{ij,a}_\al \mS^{ji,b}_{-\al} \vf_\al(z_{ba}, \om_\al + \frac{q_{ij}}{N})\,,
        }
\end{array}
\eq
\beq\label{a47}
\begin{array}{c}
 \displaystyle{
    H_{2,a} = \frac{1}{2} \sum_{i,j} \sum_\al \mS^{ij, a}_\al \mS^{ji, a}_{-\al}\,,
    }
 \end{array}
 \eq
where the function $\rho(z)$ in (\ref{H0}) is (\ref{a973}) and the function  $f_\al(z,u)$ is defined by (\ref{derphi}), (\ref{varf}). Notice that since $f(z,u)$ is a derivative of $\phi(z,u)$ with respect to the second argument, it does not have pole at $z=0$, i.e. $f(0,u)$ is well defined. It is given by (\ref{f(0,u)}).

It is easy to see that the Hamiltonians $H_{2,a}$ (\ref{a47}) are in fact the Casimir functions. They provide trivial dynamics. By fixing their levels we restrict the spin part of the phase space to the product of $n$  orbits
$\mO^1\times...\times \mO^n$ of the coadjoint action of Lie group ${\rm GL}(NM,\mC)$.

 In the spin Calogero-Moser case we had additional constraints (\ref{q0003}) generated by the action of the Cartan subgroup of ${\rm GL}(M,\mC)$. Here we deal with the $M$-dimensional Cartan subgroup $H_M\subset H\subset{\rm GL}(NM,\mC)$
 in the Cartan subgroup of ${\rm GL}(NM,\mC)$  \cite{LOSZ}. Its common action on all the orbits provides the following moment map generalizing the additional constraints (\ref{q0003}):
\beq\label{a48}\begin{array}{c}
 \displaystyle{
  \sum_{a=1}^n \mS^{kk,a}_{0,0} =\sum_{a=1}^n \mS^{ll,a}_{0,0} = {\rm const}, \quad \forall k,l=1,...,M\,.
}
\end{array}\eq
 Together with some gauge fixation conditions we come to the final description of the (spin part of the) phase space:
 $(\mO^1\times...\times \mO^n)//H_M$.
 But similarly to the spin Calogero-Moser case we do not
 perform this reduction. Instead, we will write down the Lax equations with additional unwanted term likewise we did it
 in (\ref{a35}).

 It follows from the  behaviour of $E_1(z-z_a)$ function  on the lattice $\mZ\oplus\mZ\tau$ (\ref{percond}) that the Lax matrix (\ref{nerL})
 becomes quasi-periodic when the constant in the r.h.s. of (\ref{a48}) is chosen to be zero. In this case
 the expression (\ref{a45})  is a double-periodic function of $z$ variable. Therefore, the sum of residues equals zero:
 \beq\label{a49}\begin{array}{c}
 \displaystyle{
  \sum_{a=1}^n H_{1,a}|_{(\ref{a48})\,, {\rm const}=0} = 0\,.
}
\end{array}\eq
Alternatively, one could redefine the Lax matrix by making the following shift:
 \beq\label{a491}\begin{array}{c}
 \displaystyle{
  \mL(z)\rightarrow \mL(z)-1_{NM}\sum\limits_{a=1}^n\frac{{\rm const}}{n}E_1(z-z_a)\,.
}
\end{array}\eq
On the one hand, such modified Lax matrix satisfies the same Lax equation since the scalar
non-dynamical term does not effect it. On the other hand, it is equivalent to redefinition of residues
$\mS^{ii,a}_{0,0}\rightarrow \mS^{ii,a}_{0,0}-{\rm const}/n$, so that for new set of $\mS^{ii,a}_{0,0}$
the constraints (\ref{a48}) hold with the zero r.h.s..

\paragraph{Lax pair for the flow of Hamiltonian $H_{0}$.} Consider dynamics generated by the Hamiltonian
(\ref{H0}) on the unreduced phase space $\mO^1\times...\mO^n$ with the Poisson brackets (\ref{a15}) and (\ref{a43}).
Equations of motion take the form:
\beq\begin{array}{c} \displaystyle{
\label{qp}
    \frac{d}{dt_0}{ q}_i = p_i,\quad \frac{d}{dt_0}{ p}_i = \frac{1}{N} \sum\limits_{k: k\neq i}^M \sum\limits_{a, b=1}^n \sum_{\al} \mS^{ki,a}_\al \mS^{ik,b}_{-\al} f'_\al (z_{ba}, \om_\al + \frac{q_{ki}}{N})\,,
}\end{array}\eq
$$
\displaystyle{
    \frac{d}{dt_0}{ \mS}_\al^{ij, a} = \frac{1}{N} \sum\limits_{b=1}^n \mS^{ij,a}_\al ( \mS^{jj, b}_{0,0} - \mS^{ii, b}_{0,0}) \rho(z_{b a}) + \frac{1}{N} \sum\limits_{b=1}^n \sum_{\be\neq 0} \mS^{ij, a}_{\al-\be}(\ka_{\al,\be} \mS^{jj, b}_\be - \ka_{\be, \al} \mS^{ii, b}_\be) f_\be (z_{ab}, \om_\be) + }
    $$
    \beq\label{Sij}
\begin{array}{c}
    \displaystyle{
     + \frac{1}{N} \sum\limits_{b=1}^n \sum_{\be} \Big(
     \sum\limits_{k:k\neq j}^M  \ka_{\al,\be}  \mS^{ik, a}_{\al-\be} \mS^{kj, b}_\be f_\be (z_{ab},\om_\be + \frac{q_{kj}}{N}) - \sum\limits_{k:k\neq i}^M \ka_{\be,\al}  \mS^{kj, a}_{\al-\be} \mS^{ik, b}_\be f_\be (z_{ab}, \om_\be + \frac{q_{ik}}{N})\Big)\,.
    }
\end{array}\eq
The function $f'_\al(z,u)$ entering (\ref{qp}) is a derivative of $f_\al(z,u)$ with respect to the second argument, so that
$f'_\al(z,u)=\p^2_u\vf_\al(z,u)$. It also follows from (\ref{f(0,u)}) that $f'(0,u)=-E_2'(u)=-\wp'(u)$. Let us also write down equations (\ref{Sij}) in some particular cases:
\beq\label{nerSii}\begin{array}{c}
    \displaystyle{
    \frac{d}{dt_0}{ \mS}_\al^{ii, a} = \frac{1}{N} \sum\limits_{b=1}^n \sum_{\be\neq 0} (\ka_{\al,\be} - \ka_{\be, \al}) \mS^{ii,a}_{\al-\be} \mS^{ii,b}_\be f_\be (z_{ab}, \om_\be) +}
    \\
    \displaystyle{ + \frac{1}{N} \sum\limits_{k: k\neq i}^M \sum\limits_{b=1}^n \sum_{\be} \Big( \ka_{\al,\be} \mS^{ik, a}_{\al-\be} \mS^{ki, b}_\be f_\be (z_{ab}, \om_\be + \frac{q_{ki}}{N}) - \ka_{\be,\al} \mS^{ik,b}_\be \mS^{ki, a}_{\al-\be} f_\be (z_{ab}, \om_\be + \frac{q_{ik}}{N}) \Big)\,,
    }
\end{array}
\eq
\beq\begin{array}{c}
\label{nerS00}
    \displaystyle{ \frac{d}{dt_0}{ \mS}_{0,0}^{ii, a} = \frac{1}{N} \sum\limits_{k: k\neq i}^M
    \sum\limits_{b=1}^n \sum_{\al} \Big( \mS^{ik, a}_{-\al} \mS^{ki, b}_\al  f_\al (z_{ab}, \om_\al + \frac{q_{ki}}{N}) - \mS^{ik, b}_\al \mS^{ki, a}_{-\al} f_\al (z_{ab}, \om_\al + \frac{q_{i k}}{N}) \Big)\,.
    }
\end{array}
\eq
In the case of the spin Calogero-Moser model we saw that the constraints $S_{ii}={\rm const}$ are saved by dynamics, i.e. ${\dot S}_{ii}=0$ (\ref{a34}). The same happens in the general model. To see it one should sum up equations
(\ref{nerS00}) over $a=1,...,n$:
\beq\begin{array}{c}
\label{nerS001}
    \displaystyle{
     \frac{d}{dt_0}\sum\limits_{a=1}^n{\mS}_{0,0}^{ii, a} = \frac{1}{N} \sum\limits_{k: k\neq i}^M \sum\limits_{a,b=1}^n \sum_{\al} \Big( \mS^{ik, a}_{-\al} \mS^{ki, b}_\al  f_\al (z_{ab}, \om_\al + \frac{q_{ki}}{N}) - \mS^{ik, b}_\al \mS^{ki, a}_{-\al} f_\al (z_{ab}, \om_\al + \frac{q_{ik}}{N}) \Big)\,,
    }
\end{array}
\eq
 where $t_0$ is the time of the Hamiltonian $H_0$. Expression in the r.h.s. of (\ref{nerS001}) equals zero. Indeed, by interchanging the
 summation indices $a \leftrightarrow b$ together with $\al \leftrightarrow -\al$ and using the property (\ref{derphi}) $f(-z,-u)=f(z,u)$ one easily obtains that the r.h.s is equal to itself with the opposite sign.

Introduce the $M$-matrix:
\beq\label{M00}
\begin{array}{c}
   \displaystyle{
    \mM_0(z) = \sum_{i,j=1}^M E_{ij} \otimes \mM_0^{ij}(z) \in {\rm Mat}(NM,\mC)\,, \quad \mM_0^{ij}(z) \in {\rm Mat}(N, \mC)\,,
    }
    \\
    \displaystyle{
     \mM^{ij}_0(z) = \frac{\de_{ij}}{N} \sum\limits_{a=1}^n \mS^{ii, a}_{0,0}\, 1_N\, \rho(z - z_a) + \frac{\de_{ij}}{N} \sum\limits_{a=1}^n \sum_{\al \neq 0}  \mS^{ii, a}_\al T_\al f_\al (z-z_a, \om_\al) +
     }
      \\
      \displaystyle{
       + \frac{1}{N} (1-\de_{ij}) \sum\limits_{a=1}^n \sum_\al \mS^{ij, a}_\al T_\al f_\al (z-z_a, \om_\al + \frac{q_{ij}}{N})\,.
      }
\end{array}
\eq
We come to the main statement of this subsection.
\begin{predl}
 Equations of motion (\ref{qp})-(\ref{nerS00}) are equivalent to the Lax equations with additional term:
 \beq\begin{array}{c} \label{nonLax}
    \displaystyle{ \frac{d}{dt_0}\, \mL(z) = [ \mL(z), \mM_0(z) ] -
     \frac{1}{2N} \sum\limits_{i\neq j}^M \sum\limits_{a,b=1}^n \sum_\al \mS^{ij,b}_\al 
     (\mS^{ii,a}_{0,0} - \mS^{jj,a}_{0,0}) E_{ij} \otimes T_\al f_\al' (z-z_b, \om_\al + \frac{q_{ij}}{N}).
    }
\end{array}\eq
The additional term vanishes on the constraints (\ref{a48}).
\end{predl}
The proof is straightforward though cumbersome. It is based on the usage of (\ref{derdif})-(\ref{sigma2}).

\paragraph{Lax pairs for the flows of Hamiltonians $H_{1,a}$.}

 Consider dynamics generated by the Hamiltonian $H_{1,a}$ (\ref{a46}). Similarly to the previous paragraph we assume
 the Poisson structure (\ref{a15}) and (\ref{a43}), so that the constraints (\ref{a48}) are not imposed yet.
Equations of motion are of the form (the dot below means derivative with respect to $t_{a,1}$ time variable):
\beq \displaystyle{
    \label{qpa}
    {\dot q}_i = \mS^{ii,a}_{0,0}, \quad {\dot p}_i = \frac{1}{N} \sum\limits_{k: k\neq i}^M \sum\limits_{b: b\neq a}^n
     \sum_{\al}\Big(
      \mS^{ik, a}_\al \mS^{ki, b}_{-\al} f_\al(z_{ba}, \om_\al + \frac{q_{ik}}{N}) - \mS^{ki, a}_\al \mS^{ik, b}_{-\al} f_\al(z_{ba}, \om_\al + \frac{q_{ki}}{N})\Big)\,,
    }
\eq
\beq \begin{array}{c}
\label{h1asij1}
    \displaystyle{ {\dot \mS}_\al^{ij, b} = \frac{1}{N} \mS^{ij, b}_\al (\mS^{jj, a}_{0,0} - \mS^{ii, a}_{0,0}) E_1(z_{ab}) + \frac{1}{N}
     \sum_{\be\neq 0}  \mS^{ij, b}_{\al - \be}(\ka_{\be,\al} \mS^{ii, a}_\be - \ka_{\al,\be} \mS^{jj, a}_\be) \vf_\be (z_{ba},\om_\be) + } \\
    \displaystyle{ + \frac{1}{N} \sum_{\be} \Big( \sum\limits_{k\neq i}^M \ka_{\be,\al} \mS^{kj, b}_{\al - \be} \mS^{ik, a}_\be \vf_\be (z_{ba}, \om_\be + \frac{q_{ik}}{N}) - \sum\limits_{k\neq j}^M \ka_{\al,\be} \mS^{ik, b}_{\al - \be} \mS^{kj, a}_\be \vf_\be (z_{ba}, \om_\be + \frac{q_{kj}}{N}) \Big) }
\end{array}\eq
for $b\neq a$, and
\beq \begin{array}{c}
\label{h1asij2}
    \displaystyle{ {\dot \mS}_\al^{ij, a} = -\frac{1}{N} (p_i-p_j) \mS^{ij, a}_\al   + } \\ \ \\
    \displaystyle{ +
    \frac{1}{N} \sum\limits^n_{c:\, c\neq a} \mS^{ij,a}_\al (\mS^{jj, c}_{0,0} - \mS^{ii, c}_{0,0}) E_1(z_{ac})
     + \frac{1}{N} \sum\limits^n_{c:\,c\neq a} \sum_{\be\neq 0}   \mS^{ij, a}_{\al - \be} (\ka_{\al, \be} \mS^{jj, c}_{ \be} - \ka_{\be, \al} \mS^{ii, c}_{\be})\vf_\be (z_{ac}, \om_\be) + } \\
    \displaystyle{ + \frac{1}{N} \sum\limits^n_{c:\, c\neq a} \sum_{\be}  \Big( \sum\limits_{k\neq j}^M \ka_{\al, \be} \mS^{ik, a}_{\al - \be} \mS^{kj, c}_{\be} \vf_\be (z_{ac}, \om_\be + \frac{q_{kj}}{N}) -
    \sum\limits_{k\neq i}^M \ka_{\be, \al} \mS^{kj, a}_{\al - \be} \mS^{ik, c}_{\be} \vf_\be (z_{ac}, \om_\be + \frac{q_{ik}}{N})\Big)\,.}
\end{array}
\eq
In some particular cases ($i=j$ and $\al=0$) we have
\beq\begin{array}{c}
\label{h1asii}
    \displaystyle{ {\dot\mS}_\al^{ii, b} = \frac{1}{N} \sum_{\be\neq 0} (\ka_{\be,\al} - \ka_{\al,\be}) \Big( (1 - \de_{ab})  \mS^{ii, b}_{\al - \be} \mS^{ii, a}_\be \vf_\be (z_{ba},\om_\be) - \sum_{c: c\neq a} \de_{ab} \mS^{ii, a}_{\al + \be} \mS^{ii, c}_{ -\be} \vf_\be (z_{ca},\om_\be) \Big) + }
    \\ \ \\
    \displaystyle{ + \frac{1}{N} \sum_{k: k\neq i} \sum_{\be} (1 - \de_{ab}) \Big(\ka_{\be,\al} \mS^{ki, b}_{\al - \be} \mS^{ik, a}_\be  \vf_\be (z_{ba}, \om_\be + \frac{q_{ik}}{N}) - \ka_{\al,\be} \mS^{ik, b}_{\al - \be} \mS^{ki, a}_\be \vf_\be (z_{ba}, \om_\be + \frac{q_{ki}}{N})\Big) + } \\ \ \\
    \displaystyle{ + \frac{1}{N} \sum_{k: k\neq i} \sum_{\be} \sum_{c: c\neq a} \de_{ab} \Big(\ka_{\al,\be} \mS^{ki, a}_{\al + \be} \mS^{ik, c}_{-\be} \vf_\be (z_{ca}, \om_\be + \frac{q_{ki}}{N}) -  \ka_{\be,\al} \mS^{ik, a}_{\al + \be} \mS^{ki, c}_{-\be} \vf_\be (z_{ca}, \om_\be + \frac{q_{ik}}{N})\Big)\,,}
\end{array}\eq
and
\beq\begin{array}{c}
\label{h1a00}
    \displaystyle{ {\dot\mS}_{0,0}^{ii, b} = \frac{1}{N} \sum_{k: k\neq i} \sum_{\be} (1 - \de_{ab}) \Big( \mS^{ki, b}_{- \be} \mS^{ik, a}_\be \vf_\be (z_{ba}, \om_\be + \frac{q_{ik}}{N}) - \mS^{ik, b}_{- \be} \mS^{ki, a}_\be \vf_\be (z_{ba}, \om_\be + \frac{q_{ki}}{N}) \Big) + } \\ \ \\ \displaystyle{ + \frac{1}{N} \sum_{k: k\neq i} \sum_{c: c\neq a} \sum_{\be} \de_{ab} \Big( \mS^{ki, a}_{\be} \mS^{ik, c}_{-\be} \vf_\be (z_{ca}, \om_\be + \frac{q_{ki}}{N}) - \mS^{ik, a}_{\be} \mS^{ki, c}_{-\be} \vf_\be (z_{ca}, \om_\be + \frac{q_{ik}}{N})\Big)\,,}
\end{array}\eq
where we have unified the cases $a=b$ and $a\neq b$.

Consider the $M$-matrix
\beq\begin{array}{c}
\label{M1a}
    \displaystyle{ \mM_{1,a} (z) = \sum_{i,j=1}^M E_{ij} \otimes \mM_{1,a}^{ij} (z), } \\ \ \\
    \displaystyle{ \mM_{1,a}^{ij}(z) = - \frac{\de_{ij}}{N} \mS^{ii, a}_{0,0}  1_N  E_1(z-z_a) - \frac{\de_{ij}}{N} \sum_{\al \neq 0} \mS^{ii, a}_\al T_\al \vf_\al (z-z_a, \om_\al) - } \\ \ \\
    \displaystyle{ - \frac{1}{N} (1-\de_{ij}) \sum_\al \mS^{ij, a}_\al T_\al \vf_\al (z-z_a,\om_\al +\frac{q_{ij}}{N})\,.}
\end{array}\eq
Then the following statement holds.

\begin{predl}
Equations of motion (\ref{qpa})-(\ref{h1asij2}) are equivalently written in the form of the Lax equation with
additional term
\beq\label{nonLaxh1a}\begin{array}{c}
  \displaystyle{ {\dot \mL}(z) = [ \mL(z), \mM_{1,a}(z) ] + 
  \frac{1}{N} \sum_{i\neq j}^M \sum_{b=1}^n \sum_\al \mS^{ij, a}_\al (\mS^{ii,b}_{0,0} - 
  \mS^{jj,b}_{0,0}) E_{ij} \otimes T_\al f_\al (z-z_a, \om_\al + \frac{q_{ij}}{N}) }\,,
\end{array}\eq
where the dot is the derivative with respect to time variable $t_{1,a}$. The additional term vanishes on-shell the constraints (\ref{a48}).
\end{predl}


\section{Particular cases}\label{sect4}
\setcounter{equation}{0}
\subsection*{Tops and Gaudin models}

\paragraph{${\rm gl}_N^{\times n}$ Gaudin model (case 2).}
Let us start with the second integrable family in our scheme -- the top like models. In the case $M=1$ our $gl_{NM}^{\times n}$ general model turns into elliptic Gaudin model \cite{STS}. In this model we have only the spin part of the phase space, which is now isomorphic to a direct product of $n$ orbits: $\mO_1 \times \dots \times \mO_n$. The Poisson structure (\ref{a43}) here takes form:
\beq
\{ S^a_\al, S^b_\be \} = \de^{ab} (\ka_{\al, \be} - \ka_{\be, \al}) S^a_{\al + \be}.
\eq
Here we dropped the factor $1/N$ which we introduced in (\ref{a43}).
Notice that the Poisson brackets of a diagonal scalar elements of spin $S^c_{0,0}$  with any other spin variable are equal to zero. This, together with the fact that all terms in the matrices $L$ and $M$ with this scalar diagonal spin commute with any terms ($S^c_{0,0}$ is a coefficient behind identity matrix), allows us to get rid of all such terms in the Hamiltonians and the Lax matrices. For the Lax matrix we have:
\begin{equation}
\label{BigMac}
    L(z) = \sum_a \sum_{\alpha \neq 0} S^a_\alpha T_\alpha \varphi_\alpha (z - z_a, \omega_\alpha ).
\end{equation}
Using (\ref{a45}) we get the Hamiltonians (see \cite{sigma} for the calculation details):
\beq \begin{array}{c}
\displaystyle{
    H_0 = \frac{1}{2} \sum\limits_{a,b=1}^n \sum_{\al \neq 0} S^{a}_\al S^{b}_{-\al} f_\al (z_{ba}, \om_\al), }  \\ \ \\
 \displaystyle{
    H_{1,a} = -\sum\limits_{b: b\neq a}^n \sum_{\al \neq 0} S^{a}_\al S^{b}_{-\al} \vf_\al (z_{ba}, \om_\al). }
\end{array} \eq
 The corresponding equations of motion for these Hamiltonians are as follows:
\beq
\label{fries1}
    \Dot{S}^a = \sum_{b=1}^n \sum_\be \big[ S^a, S^b_\be T_\be f_\be (z_{ab}, \om_\be) \big].
\eq
\beq \begin{array}{c}
\displaystyle{
\label{fries2}
    \Dot{S}^a =  \sum_{b: b \neq a}^n \sum_\be \big[ S^a, S^b_\be T_\be \varphi_\be (z_{ab}, \om_\be) \big],
     \quad \Dot{S}^b = -\sum_\be \big[ S^b, S^a_\be T_\be \varphi_\be (z_{ba}, \om_\be) \big], \; b\neq a }.
\end{array} \eq
The Lax equations for $L$-matrix \eqref{BigMac} and $M$-matrices:
\begin{equation}
    M_0(z) =  \sum_{a=1}^n \sum_{\alpha \neq 0} S^a_\alpha T_\alpha f_\alpha (z - z_a, \omega_\alpha ), 
    \quad M_{1,a}(z) =- \sum_{\alpha \neq 0} S^a_\alpha T_\alpha \varphi_\alpha (z - z_a, \omega_\alpha ),
\end{equation}
provide equations \eqref{fries1}-\eqref{fries2} on the constrains: $\sum_a$ tr$S^a=$ const.

\paragraph{${\rm gl}_N$ integrable top (cases 5 and 8).}
In the case of one marked point the Gaudin model turns into gl$_N$ integrable top \cite{LOZ}. The phase space of this model is (a single) coadjoint orbit of Gl$_N$ Lie group: $\mO_N$. The Lax pair takes the form
\beq \begin{array}{c} \label{McD0}
\displaystyle{
    L(z, S) = \sum_{\al \neq 0} S_\al T_\al \varphi_\al (z, \om_\al), \quad M(z, S) = \sum_{\al \neq 0} S_\al T_\al f_\al (z, \om_\al)}\,,
\end{array} \eq
and the Lax equation is equivalent to the equation of motion:
\begin{equation}
\label{McD1}
    \Dot{S} = [S, J], \quad J(S) = \sum_{\alpha \neq 0} S_\alpha T_\alpha J_\alpha, \quad S = \sum_{\alpha \neq 0} S_\alpha T_\alpha,
\end{equation}
where the inverse inertia tensor  $J$ has the components
\begin{equation}
\label{McD2}
    J_\al = - E_2 (\omega_\alpha).
\end{equation}
The Hamiltonian corresponding to equations \eqref{McD1} is:
\beq \begin{array}{c} \label{McD3}
\displaystyle{
    H^{top} = \frac{1}{2N} \hbox{tr} (S \cdot J(S)) }.
\end{array} \eq

The special case of ${\rm gl}_N$ top is the case of {\em the minimal coadjoint orbit} $\mO_N^{\hbox{\tiny{min}}}$ (box 8 on the scheme). The dimension of the orbit (and therefore, of the phase space) depends on the eigenvalues of $S$ which are fixed by the Casimir functions tr$S^k$. The case of minimal orbit corresponds to $N-1$ coincident eigenvalues, i.e. rk$S=1$, so that the dimension of the phase space is equal to
\beq
    \hbox{dim } \mO_N^{\hbox{\tiny{min}}} = 2(N-1).
\eq

\subsection*{Multispin Calogero models}
Let us consider the special case of $N=1$. In this case the Lax matrix \eqref{nerL} loses its block structure and becomes of $M \times M$ size:
\begin{equation}
    L_{ij}^G = \delta_{ij}(p_i + \sum_{a=1}^n S_{ii}^a E_1(z-z_a)) + (1-\delta_{ij}) \sum_{a=1}^n S_{ij}^a \phi (z-z_a, q_{ij}) \in {\rm Mat}(M, \mC), \quad i,j = \overline{1,M}.
\end{equation}
This $L$-operator corresponds to the multispin generalization of the Calogero-Moser model. An elliptic version of this generalization was first introduced in \cite{NN}. This model is the most general in the first family (see box 3 on the scheme). One can easily obtain the Hamiltonians of this model from the Lax operator by using standard expression \eqref{a45}:
\begin{equation}
    H_0 = \sum_{i=1}^M \frac{p_i^2}{2} + \frac{1}{2} \sum_{i=1}^M \sum_{a,b} S_{ii}^a S_{ii}^b \rho (z_{ab}) + \frac{1}{2} \sum_{i,j: i \neq j} \sum_{a,b} S_{ij}^a S_{ji}^b f(z_{ba}, q_{ij}),
\end{equation}
\begin{equation}
    H_{1,a} = \sum_{i=1}^M p_i S_{ii}^a + \sum_{i=1}^M \sum_{b: b\neq a} S_{ii}^a S_{ii}^b E_1(z_{ab}) + \sum_{i,j: i \neq j} \sum_{b: b \neq a} S_{ij}^b S_{ji}^a \phi(z_{ab}, q_{ij}),
\end{equation}
and  $H_{2,a}$ are the Casimir functions. The Poisson structure for spin variables \eqref{a43} degenerates into the Poisson-Lie structure on ${\rm gl}^*(M,\mC)$ (as for the Calogero-Moser spin model) for components of the same spin and vanishes for components of different spins:
\beq \begin{array}{c} \displaystyle{
    \{S^a_{ij}, S^b_{kl}\} = \de^{ab} ( S^a_{kj}\delta_{il} - S^a_{il}\delta_{kj} ) \,. }
\end{array} \eq
This Poisson structure provides the following equations of motion for $H_0$ Hamiltonian:
\beq \begin{array}{c} \displaystyle{
    \Dot{q}_i = p_i,\quad \Dot{p}_i = \sum_{k: k\neq i} \sum_{a, b} S^a_{ki} S^b_{ik} f' (z_{ba}, q_{ki}),} 
    \\ \ \\ 
    \displaystyle{ 
    \Dot{S}^a_{ij} = \sum\limits_{b=1}^n S^a_{ij} ( S^b_{jj} - S^b_{ii}) \rho(z_{b a}) + \sum_{b=1}^n \Big( \sum_{k: k\neq j} S^a_{ik} S^b_{kj} f (z_{ab}, q_{kj}) - \sum_{k: k\neq i} S^a_{kj} S^b_{ik} f (z_{ab}, q_{ik}) \Big)\,,
}\end{array} \eq
and for $H_{1,a}$ Hamiltonians:
\beq \begin{array}{c} \displaystyle{
    {\dot q}_i = S^{ii,a}_{0,0}, \quad {\dot p}_i = \sum\limits_{k: k\neq i} \sum\limits_{b: b\neq a} \Big( S^a_{ik} S^b_{ki} f (z_{ba}, q_{ik}) - S^b_{ik} S^a_{ki} f (z_{ba}, q_{ki}) \Big)\,, } \\ \ \\
    \displaystyle{ {\dot S}_{ij}^a = - p_{ij} S^a_{ij} + \sum\limits_{c:\, c\neq a} \Big( S^a_{ij} (S^c_{jj} - S^c_{ii}) E_1(z_{ac}) + \sum\limits_{k\neq j} S^a_{ik} S^c_{kj} \phi (z_{ac}, q_{kj}) - \sum\limits_{k\neq i} S^a_{kj} S^c_{ik} \phi (z_{ac}, q_{ik}) \Big),} \\ \ \\
    \displaystyle{ {\dot S}_{ij}^b = S^{ij, b}_\al ( S^{jj, a}_{0,0} - S^{ii, a}_{0,0}) E_1(z_{ab}) + \sum\limits_{k\neq i} S^b_{kj} S^a_{ik} \phi (z_{ba}, q_{ik}) - \sum\limits_{k\neq j} S^b_{ik} S^a_{kj} \phi (z_{ba}, q_{kj}), \; b\neq a.}
\end{array}\eq
These equations of motion are equivalent to the Lax equations with additional term:
\beq \begin{array}{c} \displaystyle{
    \frac{d}{d t_0} L(z) = [L(z), M_{0} (z) ] -
     \frac{1}{2} \sum_{i\neq j}^M \sum_{a,b=1}^n S^b_{ij} (S^a_{ii} - S^a_{jj}) E_{ij} f'(z-z_b, q_{ij}) , }
\end{array}\eq
\beq \begin{array}{c} \displaystyle{
    \frac{d}{d t_a} L(z) = [L(z), M_{1,a} (z) ] + \sum_{i\neq j}^M \sum_{b=1}^n S^a_{ij} (S^b_{ii} - S^b_{jj}) E_{ij} f(z-z_a, q_{ij}) , }
\end{array}\eq
where the corresponding $M$ matrices are of the form:
\beq \begin{array}{c} \displaystyle{
    M_0^{ij} = \delta_{ij} \sum_{a=1}^n S_{ii}^a \rho(z-z_a) + (1 - \delta_{ij}) \sum_{a=1}^n S_{ij}^a f(z-z_a, q_{ij}),}
     \\ \displaystyle{ M^{ij}_{1,a} = -\delta_{ij} S_{ii}^a E_1(z-z_a) - (1 - \delta_{ij}) S_{ij}^a \phi(z-z_a,q_{ij}).}
\end{array}\eq
The additional terms disappear under the spin constrains that in the case of the multispin Calogero-Moser model take form:
\begin{equation}
\label{MSCMrest}
    \sum_{a=1}^n S_{ii}^a =\sum_{a=1}^n S_{jj}^a = \text{const}, \; i,j=1,...,M.
\end{equation}
\subsection*{Interacting tops}
For $n=1$ the general gl$^{\times n}_{NM}$ model degenerates into elliptic gl$_{NM}$ mixed type model \cite{Polych, ZL} (see box 4 on the scheme). In this case, we have only one pole on the elliptic curve and therefore have only one type of spin variables. However, in contrast to the Calogero-Moser spin model, here the spin variables are matrix valued, which make this model a top like model.

The Lax matrix of mixed type model preserves the block structure:
\beq \begin{array}{c} \displaystyle{
    \mL(z) = \sum_{i,j=1}^M E_{ij} \otimes \mL^{ij}(z) \in {\rm Mat}(NM,\mC)\,, \quad \mL^{ij}(z) \in {\rm Mat}(N, \mC)\,, }
\end{array} \eq
where each block is defined as
\beq \begin{array}{c}
\label{cheeseburger}
    \displaystyle{ \mL^{ij}(z) = \de_{ij} \Big( p_i 1_N + S^{ii}_{0,0}\, 1_N\,  E_1(z) + \sum_{\al \neq 0} S^{ii}_\al T_\al \vf_\al (z, \om_\al) \Big) + } \\ \ \\ \displaystyle{ + (1 - \de_{ij}) \sum_\al S^{ij}_\al T_\al \vf_\al (z, \om_\al + \frac{q_{ij}}{N})\,. }
\end{array} \eq
In the case of single pole second term vanishes in the expression \eqref{a45} and we are left with the Hamiltonian
\beq \begin{array}{c}
\displaystyle{
\label{Hamburger}
    H = \sum\limits_{i=1}^M \frac{p_i^2}{2} - \frac{1}{2} \sum\limits_{i=1}^M \sum_{\al \neq 0} S^{ii}_\al S^{ii}_{-\al} E_2 (\om_\al) - \frac{1}{2} \sum\limits_{i\neq j}^M \sum_{\al} S^{ij}_\al S^{ji}_{-\al} E_2 (\om_\al + \frac{q_{ij}}{N})\,. }
\end{array} \eq
Equations of motion for this Hamiltonian are of the form:
\beq\begin{array}{c} \displaystyle{
    {\dot q}_i = p_i,\quad {\dot p}_i = - \frac{1}{N} \sum\limits_{k: k\neq i}^M \sum_{\al} \mS^{ki}_\al \mS^{ik}_{-\al} E'_2 (\om_\al + \frac{q_{ki}}{N})\,,
}\end{array}\eq
\beq\begin{array}{c} \displaystyle{
    {\dot \mS}_\al^{ij} = - \frac{1}{N} \sum_{\be\neq 0} \mS^{ij}_{\al-\be}(\ka_{\al,\be} \mS^{jj}_\be - \ka_{\be, \al} \mS^{ii}_\be) E_2 (\om_\be) + }
    \\
    \displaystyle{
    - \frac{1}{N} \sum_{\be} \Big( \sum\limits_{k:k\neq j}^M \ka_{\al,\be}  \mS^{ik}_{\al-\be} \mS^{kj}_\be E_2 (\om_\be + \frac{q_{kj}}{N}) - \sum\limits_{k:k\neq i}^M \ka_{\be,\al} \mS^{kj}_{\al-\be} \mS^{ik}_\be E_2 (\om_\be + \frac{q_{ik}}{N}) \Big)\,. }
\end{array}\eq
These equations are equivalent to the Lax equation for \eqref{cheeseburger} with the $M$-matrix
\beq \begin{array}{c}
   \displaystyle{ \mM^{ij} (z) = \frac{\de_{ij}}{N} \mS^{ii}_{0,0}\, 1_N\, \rho(z) + \frac{\de_{ij}}{N} \sum_{\al \neq 0}  \mS^{ii}_\al T_\al f_\al (z, \om_\al) + \frac{1}{N} (1-\de_{ij}) \sum_\al \mS^{ij}_\al T_\al f_\al (z, \om_\al + \frac{q_{ij}}{N})\,, }
\end{array} \eq
on the constrains \eqref{a48}, which now take form: tr$\mS^{ii} = $ const $\forall \; i$.

In special case of rk$(S) = 1$ the mixed type model turns into {\em the model of interacting tops} \cite{ZL} (box 7 on the scheme). For this model a spin part of phase space after reduction becomes isomorphic to a product of $M$ minimal coadjoint orbits:
\beq
    \mO^{\hbox{\tiny{min}}}_{NM}//H_{M} \cong \underbrace{\mO^{\hbox{\tiny{min}}}_N \times \dots \times \mO^{\hbox{\tiny{min}}}_N}_{M \; \hbox{times}}\,,\qquad H_M: {\rm diag}(h_1 1_N,...,h_M 1_N)\,.
\eq
Comparing this with the integrable tops model we see that the interacting tops model has the same phase space as $M$ tops of minimal orbit.

Let us show how the Hamiltonian \eqref{Hamburger} changes in the rank $1$ case. As it was mentioned earlier, when rk$(S) = 1$ the spin variables can be parameterized as $\mS^{ij} = \xi^i \eta^j$. Taking into account $\mS^{ij}_\al = \hbox{tr} (\mS^{ij} T_{-\al})/N$, we get:
\beq\label{r3} \begin{array}{c}
\displaystyle{
    S^{ij}_\al S^{ji}_{-\al} = \frac{ \hbox{tr} (\eta^j T_{-\al} \xi^i) \hbox{tr} (\eta^i T_{\al} \xi^j)}{N^2} = \frac{ \hbox{tr} (\xi^j \eta^j T_{-\al} \xi^i \eta^i T_{\al})}{N^2} =
 }
 \\
 \displaystyle{
    =\frac{ \hbox{tr} (S^{jj} T_{-\al} S^{ii} T_{\al})}{N^2} = \sum\limits_{\be}\frac{\kappa^2_{\al, \be} S^{jj}_\be S^{ii}_{- \be}}{N}.}
\end{array} \eq
Plugging this expression into \eqref{Hamburger}, we get the Hamiltonian of the form:
\beq \begin{array}{c}
\displaystyle{
    H^{tops} = \sum\limits_{i=1}^M \frac{p_i^2}{2} - \frac{1}{2} \sum\limits_{i=1}^M \sum_{\al \neq 0} S^{ii}_\al S^{ii}_{-\al} E_2 (\om_\al) - \frac{1}{2N} \sum\limits_{i\neq j}^M \sum_{\al, \be} \ka^2_{\al, \be} S^{jj}_\be S^{ii}_{-\be} E_2 (\om_\al + \frac{q_{ij}}{N})\,. }
\end{array} \eq
After substitution (\ref{r3}) this Hamiltonian has a clear physical interpretation. Here the first two terms describe the kinetic (and internal) energy of $M$ tops, and the last one can be interpreted as the interaction between the tops.




\section{Generalized models: description through $R$-matrices}\label{sect5}
\setcounter{equation}{0}

In this Section we construct a generalization of the Lax pairs $\mL(z)$, $\mM_0(z)$ (\eqref{nerL}, \eqref{M00}) and $\mL(z)$, $\mM_{1,a}(z)$ (\eqref{nerL}, \eqref{M1a}). This generalization is based on the $R$-matrix formulation.
The Lax pairs can be written in terms of $R$-matrix data, and the Lax equations hold due to a set of identities.
The main identity for $R$-matrix which we use is the {\em associative Yang-Baxter equation} \cite{FK}:
\beq\begin{array}{c}
\label{AYB}
    R^{z}_{12} R^{w}_{23} = R^{w}_{13} R^{z-w}_{12} + R^{w-z}_{23} R^{z}_{13}, \quad R_{ab} = R_{ab}(q_a-q_b).\,,
\end{array}\eq
where we use the standard tensor notations for $R$-matrices, which are assumed here to be in the fundamental
representation of the ${\rm GL}_N$ Lie group. Formally, a solution of (\ref{AYB}) is not a quantum $R$-matrix since the latter (by its definition) satisfies the quantum Yang-Baxter equation
\beq\begin{array}{c}
\label{qYB}
    R^{z}_{12}  R^{z}_{13} R^{z}_{23} =  R^{z}_{23} R^{z}_{13} R^{z}_{12}\,.
\end{array}\eq
The sets of solutions of (\ref{AYB}) and (\ref{qYB}) are different although they have intersection, which includes the
elliptic quantum Baxter-Belavin ${\rm GL}_N$ $R$-matrix \cite{Baxter2} (in the fundamental
representation). We briefly describe it in the Appendix. It is easy to see that in the scalar case (\ref{qYB}) is an empty condition while (\ref{AYB}) is a non-trivial functional equation -- the genus one Fay identity (\ref{Fay}).
In the general case, it can be show that solution of (\ref{AYB}) satisfying also unitarity and skew-symmetry turns out to be a true $R$-matrix, i.e. it satisfies (\ref{qYB}).

A similarity of the addition theorem for $\phi$-function and (\ref{AYB}) provides the treatment of $R$-matrix
as non-commutative analogue of elliptic Kronecker function. This leads to a set of $R$-matrix identities similar to those known for the ordinary scalar elliptic functions \cite{FK,LOZR}. We describe some of them below.

The $R$-matrix formulation of integrable tops was suggested in \cite{LOZ8} and then in \cite{LOZ2} it was shown that
(\ref{AYB}) underlies the Lax equations. In the elliptic case the $R$-matrix formulation does not provide new models but reproduce those  described in the previous Sections. At the same time such formulation allows to include into consideration the trigonometric and rational degenerations of the described models. Another application of $R$-matrix identities comes from the above mention treatment of $R$-matrix as a matrix analogue for the $\phi$-function. This leads to the $R$-matrix valued Lax pairs \cite{LOZR}. The models of this type turn out to be closely related to the models of interacting tops \cite{GSZ}. More precisely, they are  the models of interacting tops with the quantized spin part of the phase space, while the many-body degrees of freedom remain classical. This also provides applications to the long-rang spin chains
\cite{SeZ}. In this way we see that equation (\ref{AYB}) unifies the quantum and classical integrable structures.

\subsection*{R-matrix properties and identities}\label{sec:B}

The $R$-matrix has the following local expansion near $z = 0$:
\beq\begin{array}{c}
\label{serRz}
\displaystyle{
    R^{z}_{12}(x) = \frac{1}{z} 1_N \otimes 1_N + r_{12}(x) + z \: m_{12} (x) + O(z^2),
    }
\end{array}\eq
where $r_{12}$ is the classical $r$-matrix satisfying the {\em classical Yang-Baxter equation}:
\beq\begin{array}{c}
\label{CYB}
    \displaystyle{[r_{12}, r_{13}] + [r_{12}, r_{23}] + [r_{13}, r_{23}] = 0, \quad r_{12} = r_{12}(q_1-q_2),} \\ \ \\ \displaystyle{
    r_{12}(z) = \frac{1}{z} P_{12} + r^{(0)}_{12} + z r^{(1)}_{12} (x) + O(z^2).}
\end{array}\eq
In our calculations we also use degenerations of (\ref{AYB}). In particular, we use
the following identity, which can be considered as a half of the classical Yang-Baxter equation:
\beq\begin{array}{c}
    r_{12}(x) r_{13}(x+y) -  r_{23}(y) r_{12}(x) + r_{13}(x+y) r_{23} (y) = m_{12}(x) + m_{23}(y) +m_{13}(x+y).
\end{array}\eq
The latter has the following degeneration:
\beq\begin{array}{c}
    r_{12}(x) r_{13}(x) -  r^{(0)}_{23} r_{12}(x) + r_{13}(x) r_{23}^{(0)} + \partial_x r_{13}(x) P_{23} = m_{12}(x) + m_{23}(0) +m_{13}(x).
\end{array}\eq
We also need the following expression obtained by taking three consecutive limits ($z\to 0$, $q_3 \to 0$, $q_2 \to 0$) of \eqref{AYB}:
\beq\begin{array}{c}
    [r_{12}(x), m_{13}(x)] = [r^{(0)}_{23}, m_{12}(x)] + [r_{23}^{(0)}, m_{13}(x)] + [m_{23}(0), r_{12}(x)] + [\partial_x m_{12}(x), P_{23}].
\end{array}\eq

We use $R$-matrix satisfying the following set of properties:

\noindent\underline{Expansion near $x = 0$}:
\beq\begin{array}{c}
\label{serRx}
    R^{z}_{12}(x) = \frac{1}{x} P_{12} + R^{z, (0)}_{12} + x R^{z, (1)}_{12} + O(x^2)\,,
\end{array}\eq
where $P_{12}$ is the permutation operator:
\beq\begin{array}{c}
    \displaystyle{P_{12} = \frac{1}{N} \sum_\al T_\al \otimes T_{-\al}\,.}
\end{array}\eq

\noindent\underline{The Fourier symmetry}:
\beq\label{w33}\begin{array}{c}
    R^z_{12}(x) P_{12} = R^x_{12}(z).
\end{array}\eq

\noindent\underline{Unitarity}:
\beq\label{unitarity}\begin{array}{c}
    R^{z}_{12} (x) R^{z}_{21} (-x) = (\wp (z) - \wp (x)) 1_N \times 1_N.
\end{array}\eq

\noindent\underline{Skew-symmetry}:
\beq\begin{array}{c}
    R^{z}_{12}(x) = - R^{-z}_{21}(-x), \quad r_{12} (z) = -r_{21} (-z), \quad r_{12}^{(0)} = -r_{21}^{(0)}, \quad m_{12}(z) = m_{21} (-z).
\end{array}\eq
From these properties we have the following identities for the coefficients of expansions \eqref{serRz}-\eqref{CYB}, \eqref{serRx}:
\beq\begin{array}{cc}
    R^{z, (0)}_{12} = r_{12}(z) P_{12} & r_{12}^{(0)} = r_{12}^{(0)} P_{12} \\
    R^{z, (1)}_{12} = m_{12}(z) P_{12} & r_{12}^{(1)} = m_{12}^{(0)} P_{12}
\end{array}\eq
The special notation is used for the $R$-matrix derivative:
\beq\begin{array}{c}
    F^{z}_{12} (q) = \partial_q R^{z}_{12} (q).
\end{array}\eq

Other degenerations of the associative Yang-Baxter equation \eqref{AYB}:
\beq\begin{array}{c}
    R^{z-z_a}_{12}(x) R^{z-z_b, (0)}_{23} = R^{z-z_b}_{13}(x) R^{z_{ba}}_{12}(x) + R^{z_{ab}, (0)}_{23} R^{z-z_a}_{13}(x) + P_{23} F^{z-z_a}_{13}(x),
\end{array}\eq
\beq\begin{array}{c}
    R^{z-z_a,(0)}_{12} R^{z-z_b}_{23} (x) = R^{z-z_b}_{13}(x) R^{z_{ba},(0)}_{12} + R^{z_{ab}}_{23} (x) R^{z-z_a}_{13} (x) + F^{z-z_b}_{13}(x) P_{12},
\end{array}\eq
\beq\begin{array}{c}
    R^{z-z_a}_{12}(x) R^{z-z_b}_{23} (-x) = R^{z-z_b, (0)}_{13} R^{z_{ba}}_{12}(x) + R^{z_{ab}}_{23} (-x) R^{z-z_a, (0)}_{13} + F^{z_{ba}}_{32}(x) P_{13}\,.
\end{array}\eq
By differentiating the last expression with respect to $x$, we get:
\beq\begin{array}{c}
\displaystyle{
    F^{z-z_a}_{12}(x) R^{z-z_b}_{23} (-x) - R^{z-z_a}_{12}(x) F^{z-z_b}_{23} (-x) = }
     \\ \ \\ 
     \displaystyle{ = R^{z-z_b, (0)}_{13} F^{z_{ba}}_{12} (x) - F^{z_{ab}}_{23} (-x) R^{z-z_a, (0)}_{13} + \partial_x F^{z_{ba}}_{32}(x) P_{13}}\,.
\end{array}\eq
Again, by differentiating the associative Yang-Baxter equation \eqref{AYB} with respect to $q_2$ and taking the
limit $q_3 \to q_2$, we get:
\beq\begin{array}{c}
\displaystyle{
    R^{z-z_a}_{12}(x) R^{z-z_b, (1)}_{23} - F^{z-z_a}_{12}(x) R^{z-z_b, (0)}_{23} = }
     \\ \ \\ 
     \displaystyle{ = R^{z_{ab}, (1)}_{23} R^{z-z_a}_{13} (x) - R^{z-z_b}_{13} (x) F^{z_{ba}}_{12} (x) - \frac{1}{2} P_{23} \partial_x F^{z-z_a}_{13}(x)}\,,
\end{array}\eq
while in the case of the limit $q_1 \to q_2$:
\beq\begin{array}{c}
    \displaystyle{ R^{z-z_a,(0)}_{12} F^{z-z_b}_{23} (x) - R^{z-z_a,(1)}_{12} R^{z-z_b}_{23} (x) = } 
    \\ \ \\ 
    \displaystyle{ = F^{z_{ab}}_{23} (x) R^{z-z_a}_{13} (x) - R^{z-z_b}_{13} (x) R^{z_{ba}, (1)}_{12} 
    + \frac{1}{2} \partial_x F^{z-z_b}_{13}(x) P_{12}}\,.
\end{array}\eq
Finally, we assume the following $R$-matrix traces:
\beq\begin{array}{c} \label{traceR}
    \tr_1 R^z_{12} (x) = \tr_2 R^z_{12} (x) = \phi(\frac{z}{N}, x) 1_N,
     \quad \tr_1\; r_{12} (x) = E_1 (x) 1_N, \quad \tr_1 \; m_{12} (x) = \frac1N\rho(x) 1_N\,.
\end{array}\eq

\subsection*{Lax matrix and Hamiltonians}
Following \cite{LOZR,LOZ2} let us recall $R$-matrix formulation of the integrable top.
The inverse inertia tensor is of the form:
\beq \label{JR}
J(S) = \hbox{tr}_2 (m_{12}(0) S_2), \quad S_2 = 1_N \otimes S
\eq
and the corresponding Hamiltonian reads as follows:
\beq \begin{array}{c}
    H^{top} = \frac{1}{2} \hbox{tr}_{12} (m_{12} (0) S_1 S_2).
\end{array} \eq
The equations of motion \eqref{McD1} for $J$ tensor \eqref{JR} are equivalent to the Lax equation with the following Lax pair:
\beq \label{LMR}
    L(z, S) = \hbox{tr}_2 (r_{12} (z) S_2), \quad M(z, S) = \hbox{tr}_2 (m_{12} (z) S_2).
\eq
The case of the previously described elliptic gl$_N$ top model corresponds to the Baxter-Belavin $R$-matrix. Plugging $r$ and $m$ coming from this matrix into \eqref{JR}-\eqref{LMR} we get exactly expressions \eqref{McD0}-\eqref{McD3} (up to constants not included in the equations of motion)\footnote{See Appendix for details.}.

In the general case of the gl$_{NM}^{\times n}$ model, we deal with the Lax pair for the gl$_{NM}$ mixed type model introduced in \cite{GSZ}. We extend it to the case of multiple poles. The Lax matrix still has a block-matrix structure, and the size $NM \times NM$. It takes the form:
\beq \begin{array}{c} \displaystyle{
    \mL(z) = \sum_{i,j=1}^M E_{ij} \otimes \mL^{ij}(z) \in {\rm Mat}(NM,\mC)\,, \quad \mL^{ij}(z) \in {\rm Mat}(N, \mC)\,, } \\ \ \\ \displaystyle{ \mL^{ij}(z) = \de_{ij} \Big( p_i 1_N + \sum_{a=1}^n \hbox{tr}_2 (\mS^{ii,a}_2 r_{12} (z-z_a) ) \Big) + (1 - \de_{ij}) \sum_{a=1}^n \hbox{tr}_2 (\mS^{ij,a}_2 R^{z-z_a}_{12}(q_{ij}) P_{12}).
} \end{array} \eq
Here $P_{12}$ is the permutation operator in ${\rm Mat}(N, \mC)^{\otimes 2}$. In terms of the matrix basis $T_\al$ it is $P_{12} = 1/N \sum_\al T_\al \otimes T_{-\al}$.

Let us evaluate the Hamiltonians:
\beq
\begin{array}{c} \displaystyle{
    \frac{1}{2N}\, \tr \mL^2(z) = H_0 + \sum\limits_{a=1}^n ( H_{1,a} E_1(z-z_a) + H_{2,a} E_2(z-z_a) ) +
     }
     \\
     \displaystyle{
    + \frac{1}{N^2} \sum_{i,a,b} \hbox{tr} (\mS^{ii,a}) \hbox{tr} (\mS^{ii,b}) \rho(z-z_a).
}\end{array}\eq
The last term here is treated in the same way as we did for elliptic model.
We get the following expressions for the Hamiltonians:
\beq \begin{array}{c} \label{RH0}
\displaystyle{
    \mathcal H_0 = \sum_{i=1}^M \frac{p_i^2}{2} + \frac{1}{2} \sum_{i=1}^M \sum_{a,b} \hbox{tr}_{12} (\mS^{ii,a}_1 \mS^{ii,b}_2 m_{12}(z_{ab}) ) + \frac{1}{2} \sum_{i,j: i\neq j} \sum_{a,b} \hbox{tr}_{12} (\mS^{ij,a}_1 \mS^{ji,b}_2 F^{z_{ba}}_{21}(q_{ij}) P_{12}), }
\end{array} \eq
\beq \begin{array}{c}
\displaystyle{
    \mathcal H_{1,a} = \sum_{i=1}^M \frac{p_i}{N} \hbox{tr} (\mS^{ii,a}) + \frac{1}{N} \sum_{i=1}^M \sum_{b: b\neq a} \hbox{tr}_{12} (\mS^{ii,a}_1 \mS^{ii,b}_2 r_{12}(z_{ab}) ) + } \\ \ \\ \displaystyle{ + \frac{1}{N} \sum_{i,j: i\neq j} \sum_{b: b\neq a} \hbox{tr}_{12} (\mS^{ij,a}_1 \mS^{ji,b}_2 R^{z_{ab}}_{12} (q_{ji}) P_{12}),}
\end{array} \eq
\beq \begin{array}{c} \label{RH2}
\displaystyle{
    \mathcal H_{2,a} = \frac{1}{2N} \sum_{i,j} \hbox{tr}_{12} (\mS^{ij,a}_1 \mS^{ji,a}_2 P_{12}).
} \end{array} \eq
These are generalized formulae for the Hamiltonians of the gl$_{NM}^{\times n}$ model. Indeed, substituting the Belavin-Baxter $R$-matrix into expressions \eqref{RH0}-\eqref{RH2} we obtain the Hamiltonians \eqref{H0}-\eqref{a47}.

The unreduced Poisson structure for  the spin variables \eqref{a43} and the moment map \eqref{a48} remain the same, but for our purpose it is more convenient to write them in terms of Mat$(N, \mathbb C)$-valued blocks $\mS^{ij}$. Then the Poisson brackets acquire the form
\beq \begin{array}{c} \displaystyle{
\label{RPoisson}
    \{ \mS^{ij,a}_1, \mS^{km,b}_2 \} = \de^{ab} \Big( \de^{im} P_{12} \mS^{kj,a}_1 - \de^{kj} \mS^{im,a}_1 P_{12} \Big)\,.
}\end{array}\eq
As in the scalar case \eqref{a47} the second Hamiltonian (\ref{RH2}) turns out to be the Casimir function. The moment map can be represented as follows:
\beq \begin{array}{c}
\label{Rconstr}
\displaystyle{
    \hbox{tr} \Big( \; \sum_{a=1}^n \mS^{kk,a} \; \Big) = \hbox{const}=N\nu, \quad \forall k=1,...,M\,.
} \end{array} \eq

\paragraph{Lax pair for the flow of Hamiltonian $\mathcal H_{0}$.} We carry out the same procedure as for the scalar case in Section 3. Using the Poisson brackets \eqref{a15} and \eqref{RPoisson} for the unreduced spin part of the phase space we obtain the following equations of motion for $\mH_0$ flow:
\beq \begin{array}{c}
\label{eqmRq0}
\displaystyle{
    \Dot{q}_i = p_i, \quad \Dot{p}_i = - \sum_{k: k\neq i} \sum_{a,b} \hbox{tr}_{12} \Big( \mS_1^{ik,a} \mS_2^{ki,b} \partial_{q_i} F^{z_{ba}}_{21} (q_{ik}) P_{12} \Big),
} \end{array} \eq
\beq \begin{array}{c}
\label{eqmRS0}
\displaystyle{
    \Dot{\mS}^{ij,a} = \sum_{b=1}^n \mS^{ij,a} \: \hbox{tr}_2 (\mS_2^{jj,b} m_{12} (z_{ab})  ) - \sum_{b=1}^n \hbox{tr}_2 (\mS_2^{ii,b} m_{12} (z_{ab}) ) \mS^{ij,a} + } \\ \displaystyle{ + \sum_{k: k\neq j} \sum_{b=1}^n \mS^{ik,a} \hbox{tr}_2 (\mS^{kj,b}_2 F^{z_{ab}}_{12} (q_{kj}) P_{12}) - \sum_{k: k\neq i} \sum_{b=1}^n \hbox{tr}_2 (\mS^{ik,b}_2 F^{z_{ab}}_{12} (q_{ik}) P_{12}) \mS^{kj,a}.
} \end{array} \eq
 Introduce the accompany matrices
\beq \begin{array}{c} \label{mM0} \displaystyle{
    \mM_0 (z) = \sum_{i,j=1}^M E_{ij} \otimes \mM_0^{ij}(z) \in {\rm Mat}(NM,\mC)\,, \quad \mM_0^{ij}(z) \in {\rm Mat}(N, \mC)\,,} \\ \ \\ \displaystyle{ \mM_0^{ij}(z) = \de_{ij} \sum_{a=1}^n \hbox{tr}_2 (\mS^{ii,a}_2 m_{12}(z-z_a) ) + (1 - \de_{ij}) \sum_{a=1}^n \hbox{tr}_2 (\mS^{ij,a}_2 F^{z-z_a}_{12}(q_{ij}) P_{12})\,.
} \end{array} \eq
The following statement holds.
\begin{predl}
Equations of motion (\ref{eqmRq0})-(\ref{eqmRS0}) are equivalent to the Lax equations with additional term:
\beq\begin{array}{c} \label{nonLax22}
    \displaystyle{ \frac{d}{dt_0}\, \mL(z) = [ \mL(z), \mM_0(z) ] -
     \frac{1}{2} \sum\limits_{i\neq j}^M \sum\limits_{a,b=1}^n E_{ij} \otimes
      \hbox{tr}_{23} \Big( \mS^{ij,b}_2 (\mS^{ii,a}_3 - \mS^{jj,a}_3) 
      \partial_{q_{ij}} F_{12}^{z-z_b} (q_{ij}) P_{12} \Big)\,. }
\end{array}\eq
The additional term vanishes on the constraints (\ref{Rconstr}).
\end{predl}
Notice that all equations of motion \eqref{eqmRq0}-\eqref{eqmRS0} and $\mM_0$ matrix \eqref{mM0} reproduce the corresponding equations of motion for the scalar case \eqref{qp}-\eqref{Sij} and $\mM_0$ matrix \eqref{M00} in the case of the Baxter-Belavin $R$-matrix. The proof of the above statement uses the set of described above $R$-matrix identities.

\paragraph{Lax pairs for the flows of Hamiltonians $\mathcal H_{1,a}$.}
Consider the $\mH_{1,a}$ Hamiltonian flow. Again, we start from the unreduced spin part of the phase space and use the Poisson structure \eqref{a15}, \eqref{RPoisson} to obtain the following equations of motion:
\beq \begin{array}{c}
\label{eqmRq1}
\displaystyle{
    \Dot{q}_i =\frac{1}{N} \hbox{tr} (\mS^{ii,a}), \; \Dot{p}_i =
     \frac{1}{N} \sum_{k: k\neq i} \sum_{b: b\neq a} \hbox{tr}_{12}
      \Big( \mS_1^{ik,a} \mS_2^{ki,b}  F^{z_{ba}}_{21} (q_{ik}) P_{12} - 
      \mS_1^{ik,b} \mS_2^{ki,a}  F^{z_{ab}}_{21} (q_{ik}) P_{12}\Big),
} \end{array} \eq
\beq \begin{array}{c}
\label{eqmRS1a}
\displaystyle{
    \Dot{\mS}^{ij,a} = -\frac{p_{ij}}{N} \mS^{ij,a} + \frac{1}{N} \sum_{b: b\neq a} \Big( \mS^{ij,a} \: \hbox{tr}_2 (\mS_2^{jj,b} r_{12} (z_{ab}) ) - \hbox{tr}_2 (\mS_2^{ii,b} r_{12} (z_{ab}) ) \mS^{ij,a} \Big) + } \\ \ \\ \displaystyle{ + \frac{1}{N} \sum_{b: b\neq a} \Big( \sum_{k: k\neq j} \mS^{ik,a} \: \hbox{tr}_2 (\mS^{kj,b}_2 R^{z_{ab}}_{12} (q_{kj}) P_{12}) - \sum_{k: k\neq i} \hbox{tr}_2 (\mS^{ik,b}_2 R^{z_{ab}}_{12} (q_{ik}) P_{12}) \mS^{kj,a} \Big), } \\ \ \\
    \displaystyle{ \Dot{\mS}^{ij,b} = \frac{1}{N} \Big( \hbox{tr}_2 (\mS_2^{ii,a} r_{12} (z_{ba}) ) \mS^{ij,b} - \mS^{ij,b} \: \hbox{tr}_2 (\mS_2^{jj,a} r_{12} (z_{ba}) ) \Big) + } \\ \ \\ \displaystyle{ + \frac{1}{N} \Big( \sum_{k: k\neq i} \hbox{tr}_2 (\mS^{ik,a}_2 R^{z_{ba}}_{12} (q_{ik}) P_{12}) \mS^{kj,b} - \sum_{k: k\neq j} \mS^{ik,b} \: \hbox{tr}_2 (\mS^{kj,a}_2 R^{z_{ba}}_{12} (q_{kj}) P_{12}) \Big), \quad b\neq a .}
\end{array} \eq
As in the previous case these equations transform into the equations of motion for the scalar case \eqref{qpa}-\eqref{h1asij1} when we choose $R$-matrix to be the Baxter-Belavin one. The same goes for the following $\mM$ matrix:
\beq \begin{array}{c} \displaystyle{
    \mM_{1,a} (z) = \sum_{i,j=1}^M E_{ij} \otimes \mM_{1,a}^{ij}(z) \in {\rm Mat}(NM,\mC)\,, \quad \mM_{1,a}^{ij}(z) \in {\rm Mat}(N, \mC)\,,} \\ \ \\ \displaystyle{ \mM_{1,a}^{ij}(z) = - \frac{\de_{ij}}{N} \hbox{tr}_2 (\mS^{ii,a}_2 r_{12} (z-z_a) ) - \frac{1}{N} (1 - \de_{ij}) \hbox{tr}_2 (\mS^{ij,a}_2 R^{z-z_a}_{12}(q_{ij}) P_{12}).
} \end{array} \eq
Then the following statement holds.
\begin{predl}
Equations of motion (\ref{eqmRq1})-(\ref{eqmRS1a}) are equivalently written in the form of the Lax equation with
additional term:
\beq\label{nonLaxh1a22}\begin{array}{c}
  \displaystyle{ {\dot \mL}(z) = [ \mL(z), \mM_{1,a}(z) ] +
   \frac{1}{N} \sum_{i\neq j}^M \sum_{b=1}^n E_{ij} \otimes \hbox{tr}_{23} 
   \Big( \mS^{ij, a}_2 (\mS^{ii,b}_3 - \mS^{jj,b}_3) F^{z-z_a}_{12} (q_{ij}) P_{12} \Big) }\,,
\end{array}\eq
where the dot is the derivative with respect to time variable $t_{1,a}$. The additional term vanishes on-shell the constraints (\ref{Rconstr}).
\end{predl}

\section{Schlesinger systems}\label{sect6}
\setcounter{equation}{0}

The Schlesinger systems on elliptic curves \cite{LO} can be treated as non-autonomous generalization of the
Gaudin models. The positions of marked points $z_a$ and the elliptic modular parameter $\tau$ become the time variables
related to the Hamiltonians $H_a$ and $H_0$ respectively.

The Lax equations (\ref{q0001}) are replaced by the monodromy preserving equations having form of zero curvature equation. It turns out that the latter equations can be formulated in terms of the Lax pair of the corresponding Gaudin model. The phenomenon is known as the classical Painlev\'e-Calogero correspondence \cite{LO2}. Consider, for example, the Lax pair of the Calogero-Moser model (\ref{a11})-(\ref{a12}). Then the monodromy preserving equation (with $M^{\hbox{\tiny{CM}}}(z)\to M^{\hbox{\tiny{CM}}}(z)+\nu\rho(z)1_M$)
  \beq\label{w45}
  \begin{array}{c}
  \displaystyle{
 2\pi\imath\,\frac{d}{d\tau}{L}^{\hbox{\tiny{CM}}}(z)-\frac{d}{d z}M^{\hbox{\tiny{CM}}}(z)
 =[L^{\hbox{\tiny{CM}}}(z),M^{\hbox{\tiny{CM}}}(z)]
 }
 \end{array}
 \eq
is equivalent to non-autonomous equations
  \beq\label{a141}
  \begin{array}{c}
  \displaystyle{
 (2\pi\imath)^2\frac{d^2}{d\tau^2}\,{ q}_i=\nu^2\sum\limits_{k\neq
 i}\wp'(q_{ik})\,.
 }
 \end{array}
 \eq
 The derivation of the latter statement is almost the same as for the Lax equation (\ref{q0001}). The only
 additional tool is the heat equation
  \beq\label{a142}
  \begin{array}{c}
  \displaystyle{
 2\pi\imath\p_\tau\phi(z,u)=\p_z\p_u\phi(z,u)\,.
 }
 \end{array}
 \eq
 It means that the Painlev\'e-Calogero correspondence is gauge dependent phenomenon since it is based on the special choice of gauge and normalization of the Lax pair, where the matrix elements of the Lax matrix are given by $\phi$ functions. This is the case we are dealing with in this paper.

 Similarly, for the Gaudin model we have
  \beq\label{w451}
  \begin{array}{c}
  \displaystyle{
 \frac{d}{d z_a}{L}(z)-\frac{d}{d z}M_a(z)
 =[L(z),M_a(z)]\,.
 }
 \end{array}
 \eq
To summarize the Lax pairs described in the previous Sections can be straightforwardly used
for construction of the Schlesinger systems. In fact, this statement is known for a generic elliptic model related to bundles with arbitrary characteristic class \cite{LOZ14}. For example, for the most general elliptic model we have the following statement.
\begin{predl}
 Equations of motion (\ref{qp})-(\ref{nerS00}), where the dot means ${\dot x}=(2\pi\imath/N)\frac{dx}{d\tau}$, are equivalent to the monodromy preserving equations with additional term:
 \beq\begin{array}{c} \label{nonLax7}
    \displaystyle{ \frac{2\pi\imath}{N}\frac{d}{d\tau}\, \mL(z) -\p_z \mM_0(z)= [ \mL(z), \mM_0(z) ] -
    }
    \\
    \displaystyle{
    -\frac{1}{2N} \sum\limits_{i\neq j}^M 
    \sum\limits_{a,b=1}^n \sum_\al \mS^{ij,b}_\al (\mS^{ii,a}_{0,0} - 
    \mS^{jj,a}_{0,0}) E_{ij} \otimes T_\al f_\al' (z-z_b, \om_\al + \frac{q_{ij}}{N})\,.
    }
\end{array}\eq
The additional term vanishes on the constraints (\ref{a48}).
\end{predl}

\begin{predl}
Equations of motion (\ref{qpa})-(\ref{h1asij2}) are equivalently written in the form of the monodromy preserving equation with
additional term
\beq\label{nonLaxh1a7}\begin{array}{c}
  \displaystyle{ {\frac{1}{N}\frac{d}{d z_a} \mL}(z)-\p_z \mM_{1,a}(z)= [ \mL(z), \mM_{1,a}(z) ] +
   }
   \\
  \displaystyle{
   +\frac{1}{N} \sum_{i, j=1}^M \sum_{b=1}^n \sum_\al \mS^{ij, a}_\al (\mS^{ii,b}_{0,0} - \mS^{jj,b}_{0,0}) E_{ij} \otimes T_\al f_\al (z-z_a, \om_\al + \frac{q_{ij}}{N}) }\,,
\end{array}\eq
where the dot in the equations of motion is ${\dot x}=(1/N)\frac{dx}{dz_a}$. The additional term vanishes on-shell the constraints (\ref{a48}).
\end{predl}

For the models in $R$-matrix formulation all statements are the same if the following heat equation holds
  \beq\label{a143}
  \begin{array}{c}
  \displaystyle{
 2\pi\imath\p_\tau R^z_{12}(u)=N\p_z\p_uR^z_{12}(u)\,.
 }
 \end{array}
 \eq
 The latter is true for the Baxter-Belavin $R$-matrix.


\section{Appendix: Elliptic functions}\label{sec:A}
\def\theequation{A.\arabic{equation}}
\setcounter{equation}{0}


The basic element for construction of Lax pairs is the {\em Kronecker elliptic function} \cite{Weil}:
\beq\displaystyle{
\label{philimits}
    \phi(z, u) =
            \frac{\vartheta'(0) \vartheta (z + u)}{\vartheta (z) \vartheta (u)}\,,
}\eq
defined in terms of the odd Riemann theta function
\beq\begin{array}{c} \displaystyle{
     \vartheta (z)=\vartheta (z|\tau) = -\sum_{k\in \mathbb{Z}} \exp \left( \pi i \tau (k + \frac{1}{2})^2 + 2\pi i (z + \frac{1}{2}) (k + \frac{1}{2}) \right)\,.
}\end{array}\eq
 The function (\ref{philimits}) has obvious properties:
\beq\begin{array}{c} \displaystyle{
    \phi(z, u) = \phi(u, z), \quad \phi (-z, -u) = - \phi(z, u).
}\end{array}\eq
We also need the derivative $f(z,u) = \partial_u \vf(z,u)$ given by
\beq\begin{array}{c} \displaystyle{
\label{derphi}
    f(z, u) = \phi(z, u)(E_1(z + u) - E_1(u)), \quad f(-z, -u) = f(z, u)\,,
}\end{array}\eq
where
the {\em Eisenstein functions} (the first and the second) are of the form:
\beq\begin{array}{c} \displaystyle{
\label{Elimits}
    E_1(z) =
            \partial_z \ln \vartheta(z)\,,
         \quad E_2(z) = - \partial_z E_1(z) = \wp(z) - \frac{\vartheta'''(0) }{3\vartheta'(0)}\,,
}\end{array}\eq
\beq\begin{array}{c} \displaystyle{
    E_1(- z) = -E_1(z)\,, \quad E_2(-z) = E_2(z)\,.
}\end{array}\eq
 For the above functions the following {\em local expansions} near $z=0$ hold:
\beq\begin{array}{c} \displaystyle{
\label{serphi}
    \phi(z, u) = \frac{1}{z} + E_1 (u) + z\rho (u) + O(z^2)\,,
}\end{array}\eq
\beq\begin{array}{c} \displaystyle{
\label{serE}
    E_1(z) = \frac{1}{z} + \frac{z}{3} \frac{\vartheta'''(0) }{\vartheta'(0)} + O(z^3)\,,
}\end{array}\eq
\beq\begin{array}{c} \displaystyle{
\label{f(0,u)}
    f(0, u) = - E_2(u)\,,
}\end{array}\eq
where in (\ref{serphi}) and in what follows we use notation:
\beq\label{a973}\begin{array}{c} \displaystyle{
    \rho (z) = \frac{E^2_1(z) - \wp(z)}{2}\,.
}\end{array}\eq

\noindent {\em The quasi-periodic behaviour} (on the lattice of periods $1$ and $\tau$):
\beq\begin{array}{c}
\label{percond}
    \displaystyle{ E_1 (z + 1) = E_1 (z), \quad E_1(z + \tau) = E_1(z) - 2\pi i, } \\ \ \\
    \displaystyle{ E_2 (z + 1) = E_2 (z), \quad E_2(z + \tau) = E_2(z),} \\ \ \\
    \displaystyle{ \phi (z + 1, u) = \phi (z, u), \quad \phi(z + \tau, u) = e^{- 2\pi i u} \phi(z, u), } \\ \ \\
    \displaystyle{ f (z + 1, u) = f (z, u), \quad f (z + \tau, u) = e^{-2\pi i u}(f(z, u) - 2\pi i \phi (z, u)).}
\end{array}\eq

\noindent {\em Addition formula and its degenerations}:
%
\beq \begin{array}{c} \label{Fay} \displaystyle{
    \phi(z_1, u_1) \phi(z_2, u_2) = \phi(z_1, u_1 + u_2) \phi(z_2 - z_1, u_2) + \phi(z_2, u_1 + u_2) \phi(z_1 - z_2, u_1).
}\end{array}\eq
\beq\label{derdif}\begin{array}{c} \displaystyle{
    f(z_1, u_1) \phi(z_2, u_2) - \phi(z_1, u_1) f(z_2, u_2) = \phi(z_2, u_1 + u_2) f(z_{12}, u_1) - \phi(z_1, u_1 + u_2) f(z_{21}, u_2),
}\end{array}\eq
\beq\label{a974}\begin{array}{c} \displaystyle{
    f(z, u_1) \phi(z, u_2) - \phi(z, u_1) f(z, u_2) = \phi(z, u_1 + u_2) (E_2(u_2) - E_2 (u_1)),
}\end{array}\eq
\beq\begin{array}{c} \displaystyle{
\label{diffsign}
    \phi(z, u) \phi(z, -u) = E_2(z) - E_2(u) = \wp (z) - \wp (u),
}\end{array}\eq
\beq\label{a975}\begin{array}{c} \displaystyle{
    \phi(z, u_1) \phi(z, u_2) = \phi(z, u_1 + u_2) (E_1 (z) + E_1 (u_1) + E_1 (u_2) - E_1 (z + u_1 + u_2)),
}\end{array}\eq
\beq\label{a976}\begin{array}{c} \displaystyle{
    \phi(z_1, u) \phi(z_2, u) = \phi(z_1 + z_2, u) (E_1 (z_1) + E_1 (z_2)) - f(z_1 + z_2, u).
}\end{array}\eq
\beq\begin{array}{c} \displaystyle{
\label{sigma1}
    \phi (z_1, u) \rho(z_2) - E_1 (z_2) f(z_1, u) + \phi(z_2, u) f(z_{12}, u) - \phi(z_1, u) \rho(z_{21}) = -\frac{1}{2} \partial_u f(z_1, u) ,
}\end{array}\eq
\beq\begin{array}{c} \displaystyle{
\label{Ep}
    (E_1(u+v) - E_1(u) - E_1(v))^2 = \wp(u+v) +  \wp(u) + \wp(v),
}\end{array}\eq
\beq\begin{array}{c} \displaystyle{
\label{sigma2}
    \phi (z, u) \rho(z) - E_1 (z) f(z, u) - \phi(z, u) \wp(u) = -\frac{1}{2} \partial_u f(z, u).
}\end{array}\eq

\noindent Using the Kronecker elliptic function and its derivative we define the following {\em set of functions}:
\beq\begin{array}{c} \displaystyle{
\label{varphi}
 \displaystyle{
    \vf_\al (z, \om_\al + u) = \exp (2 \pi i \frac{\al_2}{N} z) \phi (z, \om_\al + u), \quad \om_\al = \frac{\al_1 + \al_2 \tau}{N},
    }
}\end{array}\eq
\beq\begin{array}{c} \displaystyle{
\label{varf}
 \displaystyle{
    f_\al (z, \om_\al + u) = \exp (2 \pi i \frac{\al_2}{N} z) f (z, \om_\al + u),
    }
}\end{array}\eq
\beq\begin{array}{c} \displaystyle{
    \label{formulaf}
     \displaystyle{
    f_\al (z, \om_\al + u) = \partial_u \vf_\al (z, \om_\al + u) = \vf_\al (z, \om_\al + u) (E_1(z + \om_\al + u) - E_1(\om_\al + u)).}
}\end{array}\eq
The functions (\ref{varphi}) are elements of a basis in the space of sections of the ${\rm End}(V)$ for a holomorphic vector bundle $V$ (over elliptic curve) of degree 1.

\noindent The {\em addition formulae for the basis functions} take the form:
\beq\begin{array}{c}
\label{formulaPhi}
 \displaystyle{
        \vf_\al (z_1, \om_\al + u_1) \vf_\be (z_2, \om_\be + u_2) = \vf_\al (z_1 - z_2, \om_\al + u_1) \vf_{\al + \be} (z_2, \om_{\al+\be} + u_1 + u_2) +}
         \\ \ \\
          \displaystyle{
        + \vf_\be (z_2 - z_1, \om_\be + u_2) \vf_{\al + \be} (z_1, \om_{\al + \be} + u_1 + u_2)\,.
        }
\end{array}\eq
 In particular,
\beq\begin{array}{c}
\label{sinfei}
 \displaystyle{
    \vf_\al(z-z_a, \om_\al) \vf_\be(z-z_b, \om_\be) =
  }
  \\ \ \\
   \displaystyle{
    =\vf_\al(z_{ba}, \om_\al) \vf_{\al + \be}(z - z_b, \om_{\al + \be}) + \vf_\be(z_{ab}, \om_\be) \vf_{\al + \be}(z-z_a, \om_{\al + \be})
    },
\end{array}\eq
and
\beq\begin{array}{c}
\label{formulaE}
\displaystyle{
        \vf_\al (z, \om_\al + u_1) \vf_\be (z, \om_\be + u_2) = \vf_{\al + \be}(z, \om_{\al + \be} + u_1 + u_2)  \times
         }
         \\ \ \\
         \displaystyle{
         \times\Big( E_1(z)+ E_1(\om_\al + u_1) + E_1(\om_\be + u_2) - E_1(z + \om_{\al + \be} + u_1 + u_2)\Big)\,,
         }
\end{array}\eq
and
\beq\begin{array}{c}
\label{formulaEf}
\displaystyle{
        \vf_\al (z_1, \om_\al + u) \vf_\al (z_2, \om_\al + u) =
        }
        \\ \ \\
         \displaystyle{
         = \vf_{\al}(z_1 + z_2, \om_\al + u)  ( E_1(z_1) + E_1(z_2)) - f_\al (z_1 + z_2, \om_\al + u)\,.
         }
\end{array}\eq

\paragraph{The Baxter-Belavin elliptic $R$-matrix} \cite{Baxter2}:
\beq\begin{array}{c}
    \displaystyle{R^{BB}_{12} (z, x) = \sum_\al \varphi_\al (x, z + \om_\al) T_\al \otimes T_{-\al}}\in {\rm Mat}(N, \mC)^{\otimes 2}
\end{array}\eq
satisfies all required properties \eqref{serRz}-\eqref{traceR}, but with a different normalization. We use a slightly different $R$-matrix to fulfill all properties including the normalization:
\beq\begin{array}{c}
    \displaystyle{R^{z}_{12} (x) = \frac{1}{N}
    R^{BB}_{12} (z/N, x) = \frac{1}{N} \sum_\al \varphi_\al (x, \frac{z}{N} + \om_\al) T_\al \otimes T_{-\al}}.
\end{array}\eq
Using \eqref{serphi} and \eqref{serRz} we get the corresponding classical $r$-matrix and $m$-matrix:
\beq\begin{array}{c}
    \displaystyle{ r_{12} (z) = \frac{1}{N} E_1(z) 1_N \otimes 1_N + \frac{1}{N} \sum_{\al \neq 0} \varphi_\al (z, \om_\al) T_\al \otimes T_{-\al}},
\end{array}\eq
\beq\begin{array}{c}
    \displaystyle{ m_{12} (z) = \frac{1}{N^2} \rho(z) 1_N \otimes 1_N + \frac{1}{N^2} \sum_{\al \neq 0} f_\al (z, \om_\al) T_\al \otimes T_{-\al}}.
\end{array}\eq
Taking the derivative of the $r$-matrix, we get:
\beq\begin{array}{c}
    \displaystyle{F^0_{12} (z) = \partial_z r_{12}(z) = - \frac{1}{N} E_2(z) 1_N \otimes 1_N +} 
    \\ \ \\ 
    \displaystyle{ + \frac{1}{N} 
    \sum_{\al \neq 0} \varphi_\al (z, \om_\al) (E_1(z+\om_\al) - E_1(z) +2\pi i \partial_\tau \om_\al) T_\al \otimes T_{-\al}}.
\end{array}\eq
The following identity for elliptic functions (finite Fourier transformation) is useful for the Fourier symmetry
(\ref{w33}) and other applications:
\beq\begin{array}{c}
    \displaystyle{\frac{1}{N} \sum_\al \kappa^2_{\al, \be} \varphi_\al (Nx, \om_\al + \frac{z}{N}) = \varphi_\be (z, \om_\be + x), \quad \forall \be \in \mZ_N \times \mZ_N}.
\end{array}\eq
Its special cases are:
\beq\begin{array}{c}
    \displaystyle{\sum_\al E_2 (\omega_\al + x) = N^2 E_2 (N x)}
\end{array}\eq
and
\beq\begin{array}{c}
    \displaystyle{ \sum_{\al\neq(0,0)} 
    \kappa^2_{\al, \be} \varphi_\al (x, \om_\al) (E_1(x+\om_\al) - E_1(x) + 2\pi i \partial_\tau \om_\al) - E_2(x) = - E_2 (\om_\be + \frac{x}{N})}\,.
\end{array}\eq


\subsection*{Acknowledgments}
This work is supported by the Russian Science Foundation under grant
19-11-00062 and performed in Steklov Mathematical Institute of Russian Academy of Sciences.


\begin{small}
 
\end{small}


\begin{thebibliography}{99}
\addcontentsline{toc}{section}{References}


\bibitem{Atiyah} M.F. Atiyah,
Proceedings of the London Mathematical Society, s3-7 (1957)
414–-452.

\bibitem{Baxter2}
R.J. Baxter,
Ann. Phys. 76 (1973) 25--47.

L. Takhtajan, L. Faddeev,
 Russ. Math. Surveys, 34:5 (1979) 11--68.

A.A. Belavin,
Nucl. Phys. B, 180 (1981) 189--200.

M.P. Richey, C.A. Tracy, 
J. Stat. Phys., 42 (1986) 311--348.

\bibitem{BAB}
E. Billey, J. Avan, O. Babelon,
Physics
Letters A, 186 (1994) 114--118; hep-th/9312042.



\bibitem{Calogero2}
%
F. Calogero,
  Lett.
Nuovo Cim. 13 (1975) 411-–416.

F. Calogero,
 Lett. Nuovo Cim. 16 (1976) 77-–80.

J. Moser,
 Adv. Math. 16 (1975) 1-–23.

M.A. Olshanetsky, A.M. Perelomov,
 Phys. Rep. 71 (1981) 313–-400.


\bibitem{FK} S. Fomin, A.N. Kirillov,
Advances in geometry; Prog. in Mathematics book series, 172
 (1999) 147--182.

 A. Polishchuk, 
Advances in Mathematics 168:1 (2002)  56–-95.


\bibitem{Gaudin}
M. Gaudin,
J. Physique, 37:10 (1976) 1087–-1098.

M. Gaudin, La Fonction d'Onde de Bethe, Masson, Paris (1983).

   \bibitem{GH}
J. Gibbons, T. Hermsen,
Physica D: Nonlinear Phenomena, 11 (1984) 337--348;

S. Wojciechowski,
 Physics Letters A, 111 (1985) 101--103.


\bibitem{GSZ} A. Grekov, I. Sechin, A. Zotov, 
JHEP, 2019:10 (2019) 81; arXiv:1905.07820 [math-ph].

I.A. Sechin, A.V. Zotov,
 Russian Math. Surveys, 74:4 (2019) 767--769; 	arXiv:1905.08724 [math.QA].

A. Grekov, A. Zotov,
J. Phys. A, 51 (2018), 315202 , 26 pp., arXiv: 1801.00245 [math-ph].

\bibitem{Inoz} V.I. Inozemtsev,
Commun. Math. Phys. 121 (1989) 629--638 .

A.V. Zotov, Yu.B. Chernyakov,
Theoret. and Math. Phys. 129:2 (2001) 1526--1542; 	arXiv:hep-th/0102069.


 \bibitem{Hitchin1} N. Hitchin, 
Duke Math.
J. 54 (1987) 91--114.


\bibitem{Kr}
I.M.  Krichever, 
Funct.  Anal.  Appl., 14:4 (1980) 282--290; arXiv:hep-th/0108110.

\bibitem{LO} A.M. Levin, M.A. Olshanetsky,
Amer. Math. Soc. Transl. (2) 191 (1999) 223-262; arXiv:hep-th/9709207.

K. Takasaki,
Lett. Math. Phys. 44 (1998) 143-156; arXiv:hep-th/9711058.


Yu. Chernyakov, A. M. Levin, M. Olshanetsky, A. Zotov,
J. Phys. A: Math. Gen., 39:39 (2006), 12083–12101; arXiv: nlin/0602043.

\bibitem{LO2} A.M. Levin, M.A. Olshanetsky,
CRM Series in Mathematical Physics book series,
Calogero—Moser— Sutherland Models (2000) 313-332;
arXiv:alg-geom/9706010.

\bibitem{LOZ} A. Levin, M. Olshanetsky, A. Zotov,
Commun. Math. Phys.  236 (2003) 93--133;     arXiv:nlin/0110045.

A.V. Zotov, Physics of Particles and Nuclei, 37 (2006) 400-–443.

A.V. Zotov, A.V. Smirnov,
Theoret. and Math. Phys., 177:1 (2013) 1281–-1338.



\bibitem{LOSZ} A. Levin, M. Olshanetsky, A. Smirnov, A. Zotov,
Commun. Math. Phys., 316 (2012) 1--44;  arXiv:1006.0702.

A. Levin, M. Olshanetsky, A. Smirnov, A. Zotov,
 J. Geom. Phys., 62:8 (2012) 1810--1850;  arXiv:1007.4127.

 A. Levin, M. Olshanetsky, A. Smirnov, A. Zotov,
SIGMA 8 (2012) 095, 37; arXiv:1207.4386 [math-ph].


\bibitem{LOZ8} A. Levin, M. Olshanetsky, A. Zotov,
JHEP 07 (2014) 012;
arXiv:1405.7523 [hep-th].

T. Krasnov, A. Zotov, Annales Henri Poincare, 20:8 (2019)
2671--2697;
 arXiv:1812.04209 [math-ph].

 G. Aminov, S. Arthamonov, A. Smirnov, A. Zotov,
  J. Phys. A: Math. Theor., 47:30 (2014), 305207;\\ arXiv: 1402.3189 [math-ph].

\bibitem{LOZR} A. Levin, M. Olshanetsky, A. Zotov,
JHEP 10 (2014) 109; arXiv:1408.6246 [hep-th].
\\
A.M. Levin, M.A. Olshanetsky, A.V. Zotov,
Theoret. and Math. Phys. 184:1 (2015) 924--939;\\
 arXiv:1501.07351 [math-ph].

\bibitem{LOZ2} A. Levin, M. Olshanetsky, A. Zotov,
J. Phys. A: Math. Theor. 49:39 (2016) 395202; arXiv:1603.06101 [math-ph].

\bibitem{LOZ14}  A.M. Levin, M.A. Olshanetsky, A.V. Zotov,
Russian Math. Surveys, 69:1 (2014) 35–118; arXiv: 1311.4498 [math-ph].

\bibitem{NN}
N.  Nekrasov, 
 Commun.  Math.  Phys.  180 (1996) 587--604; hep-th/9503157.


\bibitem{Pasquier} V. Pasquier,
Commun. Math. Phys. 118 (1988) 355--364.

\bibitem{Polych} A.P. Polychronakos,
Phys. Rev. Lett. 89 (2002) 126403; hep-th/0112141.

A.P. Polychronakos,
 Nucl. Phys. B543 (1999) 485--498; hep-th/9810211.

A.P. Polychronakos,
J. Phys. A: Math. Gen. 39 (2006) 12793; hep-th/0607033.


\bibitem{STS} A.G. Reiman, M.A. Semenov-Tian-Shansky,
Zap. Nauchn. Sem. LOMI, 150 (1986) 104–-118;

A.G. Reiman, M.A. Semenov-Tian-Shansky, Journal of Soviet Mathematics, 46 (1989) 1631–-1640.


\bibitem{SeZ} I. Sechin, A. Zotov,
Phys. Lett. B, 781 (2018) 1–7 , arXiv: 1801.08908 [math-ph].

\bibitem{Skl} E.K. Sklyanin,
Preprint LOMI, E-3-79. Leningrad (1979).

E.K. Sklyanin,
Journal of Soviet Mathematics, 46 (1989) 1664–-1683.

L.D. Faddeev, L.A. Takhtajan,
{\em Hamiltonian methods in the theory of solitons},
Springer-Verlag, (1987).

E.K. Sklyanin,
Funct. Anal. Appl. 16:4 (1982) 263--270.


\bibitem{Weil} A. Weil, {\em Elliptic functions according to Eisenstein and
Kronecker}, Springer-Verlag, 
 (1976).

 D. Mumford, {\em Tata Lectures on Theta I, II},
Birkh\"auser, Boston, Mass. (1983, 1984).


\bibitem{ZL} A.V. Zotov, A.M. Levin,
Theoret. and Math. Phys., 146:1 (2006) 45--52.

A. Levin, M. Olshanetsky, A. Smirnov, A. Zotov,
J. Phys. A: Math. Theor. 46:3 (2013) 035201;\\ arXiv:1208.5750 [math-ph].

\bibitem{sigma} A.V. Zotov, 
SIGMA 7 (2011) 067; arXiv:1012.1072 [math-ph].








\end{thebibliography}
\end{document}